\documentclass[%
longbibliography,
reprint,
superscriptaddress,
amsmath,amssymb,
aps,
prx,
floatfix,
]{revtex4-2}
\usepackage{float}
\usepackage{graphicx}
\usepackage{dcolumn}
\usepackage{bm}

\usepackage{tikz}
\usetikzlibrary{arrows} 
\usepackage[normalem]{ulem}

\definecolor{mydarkblue}{rgb}{0.0, 0.0, 0.5}
\usepackage[colorlinks=true,
            linkcolor=mydarkblue,
            citecolor=mydarkblue,
            urlcolor=mydarkblue]{hyperref}
\usepackage{upgreek}

\newif\ifshowdeleted
\showdeletedfalse   

\begin{document}

\preprint{}

\title{Optimizing information transmission in optogenetic Wnt signaling}

\author{Olivier Witteveen}
\affiliation{
Department of Bionanoscience, Kavli Institute of Nanoscience Delft, Technische Universiteit Delft,\\
Van der Maasweg 9, 2629 HZ Delft, The Netherlands
}

\author{Samuel J. Rosen}
\affiliation{Interdisciplinary Program in Quantitative Biosciences, University of California, Santa Barbara, California 93106, USA}
\author{Ryan S. Lach}
\affiliation{Integrated Biosciences, Inc., Redwood, California 94065, USA}
\author{Maxwell Z. Wilson}
\email{mzw@ucsb.edu}
\affiliation{Department of Molecular, Cellular, and Developmental Biology, University of California, Santa Barbara, California 93106, USA}
\affiliation{Neuroscience Research Institute, University of California, Santa Barbara, California 93106, USA}
\affiliation{Biomolecular Science and Engineering, University of California, Santa Barbara, California 93106, USA}
\affiliation{Center for BioEngineering, University of California, Santa Barbara, California 93106, USA}

\author{Marianne Bauer}
\email{M.S.Bauer@tudelft.nl}
\affiliation{
Department of Bionanoscience, Kavli Institute of Nanoscience Delft, Technische Universiteit Delft,\\
Van der Maasweg 9, 2629 HZ Delft, The Netherlands
}

\date{\today}

\begin{abstract} 
Populations of cells regulate gene expression in response to external signals, but their ability to make reliable collective decisions is limited by both intrinsic noise in molecular signaling and variability between individual cells. In this work, we use optogenetic control of the canonical Wnt pathway as an example to study how reliably information about an external signal is transmitted to a population of cells, and determine an optimal encoding strategy to maximize information transmission from Wnt signals to gene expression. 
We find that it is possible to reach an information capacity beyond 1 bit only through an appropriate, discrete encoding of signals: using either no Wnt, a short Wnt pulse, or a sustained Wnt signal.
By averaging over an increasing number of outputs, we systematically vary the effective noise in the pathway. As the effective noise decreases, the optimal encoding comprises more discrete input signals.
These signals do not need to be fine-tuned to achieve near-optimal information transmission.
The optimal code transitions into a continuous code in the small-noise limit, which can be shown to be consistent with the Jeffreys prior.
We visualize the performance of different signal encodings using decoding maps. 
Our results suggest optogenetic Wnt signaling allows for regulatory control beyond a simple binary switch, and provides a framework to apply ideas from information processing to single-cell \textit{in vitro} experiments.

\end{abstract}

\maketitle

\section{Introduction}\label{sec:intro}

\begin{figure*}[!hbt]
    \centering
    \includegraphics[width=0.95\textwidth]{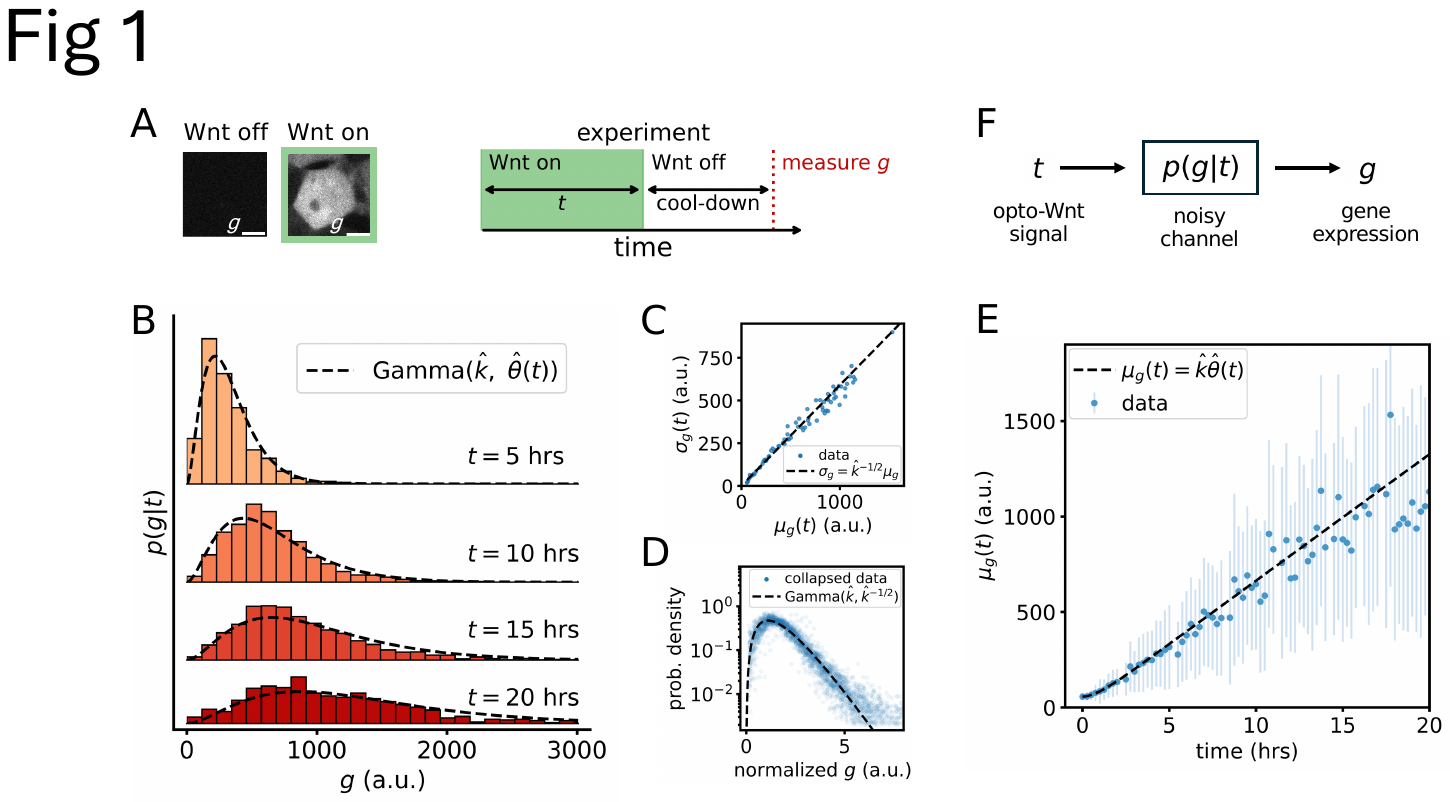}
    \caption{(A) Optogenetic control of Wnt-signaling. In the absence of light, there is no Wnt signal and no expression of TopFlash $g$. When the light is activated, Wnt target genes are expressed. We vary the duration $t$ of the Wnt signal and measure the resulting gene expression $g$. (B) The histograms over $g$ are long-tailed, left-skewed, and unimodal; shown here for Wnt signal durations $t = 5$, $10$, $15$, and $20$ hours. Black lines show the gamma distribution from Eq. \ref{eq:likelihood}, evaluated at the appropriate $t$. (C) The mean $\mu_g(t)$ of each histogram scales linearly with the standard deviation $\sigma_g(t)$. The black line shows the linear dependence predicted by Eq. \ref{eq:likelihood}. (D) Rescaling the histograms (dividing by the standard deviation) shows a collapse of gene expression data. The collapsed data is described well by a gamma distribution with shape parameter $\hat{k} \approx 2.88 \pm 0.01$ and unit variance (black line). (E) The mean gene expression $\mu_g(t)$ grows linearly with time, as captured by Eq. \ref{eq:likelihood} (black line). Error bars show the standard deviation. 
    (F) We can view our system analogously to a communication channel where input $t$ is mapped to output $g$ via the noisy transmission probability $p(g \vert t)$. }\label{fig:stdev_and_mean_tf}
\end{figure*}

Cells respond to external signals by adapting their gene expression \cite{jacob1961genetic}. 
However, gene regulatory responses can fluctuate \cite{elowitz_stochastic_2002, elf_near-critical_2003, paulsson_summing_2004, salman_universal_2012}. Especially in the context of development, precise responses are important for coordinated cell fate decisions that lead to the healthy development of an organism \cite{dubuis_positional_2013, merle_precise_2024, warmflash_method_2014}; therefore, the considerable cell-to-cell variability that has been observed in the downstream targets of signaling pathways crucial for development may seem surprising \cite{stanoev2021robustness, suderman, elAzhar}. 

The mutual information between a signal and an output quantifies the precision in information transfer \cite{cover_elements_2012, bialek_biophysics_2012}. Experiments on mammalian signaling pathways often report values barely exceeding one bit \cite{cheong_information_2011, uda_robustness_2013}, which is the minimum amount of information required for a reliable decision between only two states; for example, a differentiated and an undifferentiated state.
To understand if cells actually have more information available, a variety of approaches have been proposed in different biological contexts, including specific computational strategies, signaling architectures, cell-to-cell contact adjustments, or
incorporating the information contained in temporal dynamics \cite{suderman, iyer2023cellular, stanoev2021robustness, bettoni_optimizing_2024, selimkhanov_accurate_2014, tostevin_mutual_2009, reinhardt_path_2023, moor2023dynamic}. Here, we turn to an optimization over the space of possible input signals to investigate how information transmission in an important signaling pathway can be made more precise.  

We focus on the canonical Wnt signaling pathway, a key regulator of cell fate decisions during development and maintenance of adult tissues. Wnt signaling is crucial for the differentiation of stem cells into lineages such as skin, bone, and other tissues \cite{cadigan_wnt_1997, arnold_wnt_2019}. We study cellular responses using an established reporter of Wnt transcriptional activity, TopFlash, following optogenetic activation of the canonical Wnt signaling pathway \cite{lach_nucleation_2022, rosen_anti-resonance_2025}.
We vary the duration of the optogenetic Wnt signals and observe long-tailed distributions of gene expression; we describe them using gamma distributions. Similar distributions have been observed more widely for protein fluctuations among cells in a population and are consistent with stochastic gene expression models \cite{salman_universal_2012, friedman_linking_2006, cai_stochastic_2006}.

We optimize information transmission between optogenetic inputs and gene expression outputs by determining the optimal prior distribution over input signals.
We find that if signals are chosen uniformly from all possible signal durations, the mutual information is
approximately 0.7 bits,
whereas with a prior that involves only two discrete signals---no Wnt, or a long sustained Wnt signal---we can reach almost one bit.
We can obtain more than one bit when we fully optimize the distribution of input signals, and find that the optimal prior consists of three different input signals;
a Wnt off state, a short Wnt pulse, and a long sustained Wnt signal.
Since different maps between input and output can have the same mutual information, we visualize which input signals can be distinguished based on the output using decoding maps, for which we provide analytic expressions whenever possible.

It is possible that cells have more information available than observed with our fluorescent reporter. This increased precision could be a consequence of different biological scenarios: for example, if cells average expressed proteins among neighbors \cite{mikels_wnts_2006, erdmann2009role, gasparski_mechanoreception_2015}, or if they derive differentiation outcomes in response to not one but multiple representative Wnt targets \cite{iyer2023cellular, funa2015beta, kim_two-element_2017, kicheva2023control}. 
Therefore, we explore how these optimal prior signals change if gene expression responses are represented by those we observe, but more precise; mathematically, we describe this increased precision by a narrowing gamma distribution. We use the Blahut--Arimoto algorithm to optimize the input distribution. A central result is that the optimal input signals are discrete for the noisy gene expression we observe and smoothly transition to a continuous encoding as the noise decreases.
We show that the continuous distribution in the small-noise limit is equivalent to the Jeffreys prior. Finally, we present calculations to show that the discrete optimal priors yield `sloppy' optima \cite{gutenkunst, transtrum2015perspective}: this means that they do not need to be fine-tuned.

This signal-level optimization has potential application in engineering contexts: our optimized signals correspond to those that external users should supply such that the output can be decoded with high precision. It may also offer insight into the signals that cells might encounter in natural contexts: ideas from efficient coding  \cite{barlow_possible_1961} suggest that signal transmission is optimized, and therefore would predict that the signals we calculate should be those typically encountered by cells (see Refs. \cite{laughlin_simple_1981, bialek_reading_1991, brenner_adaptive_2000} for work on neurons).

While it is unclear if cells need to decode Wnt signal durations, how exactly Wnt provides information is also not yet established: hypotheses in different contexts include Wnt timing or duration \cite{Anton2007, spence_directed_2011, sato_growing_2013, van_den_brink_single-cell_2020}, fold-changes \cite{goentoro_evidence_2009, goentoro_incoherent_2009}, or absolute concentrations, either in gradients or dynamics \cite{sato_growing_2013,aulehla_2008, cimetta_microfluidic_2010, sonnen_modulation_2018, cooper_transport_2024}. Timing and duration of signals can play key role in guiding differentiation \cite{zon_loss_2024, rosen_anti-resonance_2025, YadavKoseska}, and we chose to investigate the duration of Wnt here because it is easily accessible in the opto-Wnt experiments. Since our work depends only on the parameterization of the conditional output distribution, our results will apply also to other gene regulatory outputs and input signals. We discuss how cells may be able to interpret signals with discrete priors inspired by recent work that optimizes an understanding or model of the world with finite samples \cite{mattingly_maximizing_2018}.

More broadly, we present a systematic framework for investigating how heterogeneity in gene expression across a population impacts information transmission, including cases where the realistic biological noise is difficult to characterize.

\section{Cellular responses to Wnt signaling}\label{sec:opto_wnt}

We explore the expression of genes that respond to the canonical Wnt signaling pathway in a clonal established human embryonic kidney cell line (HEK293T) engineered to respond to optogenetic Wnt signals \cite{lach_nucleation_2022, rosen_anti-resonance_2025}. The duration of the Wnt signal can be varied experimentally, and we use this duration $t$ as an input signal. 
We measure cellular responses to Wnt signaling using a synthetic fluorescent TCF/LEF (TopFlash-type) iRFP reporter that reflects the activation of Wnt/$\upbeta$-catenin target genes \cite{lach_nucleation_2022}. This reporter is established for Wnt targets \cite{ferrer2010sensitive, ogamino2024dynamics, fuerer2010lentiviral,lach_nucleation_2022} and we refer to it as \textit{TopFlash} from now on. At the molecular level, TopFlash and many canonical Wnt/$\upbeta$-catenin target genes are activated as a result of $\upbeta$-catenin accumulation in the cytoplasm and nucleus, following the binding of extracellular Wnt ligands to membrane receptors \cite{liu_wnt-catenin_2022}.

We collect the output expression levels (fluorescent intensity per cell) of TopFlash, denoted by $g$, of ca. $1500 \pm 800$ cells to optogenetic Wnt input signals of varying durations $t$ ranging from 0 to 20 hours (Fig. \ref{fig:stdev_and_mean_tf}A).
The experiment is conducted using a high-throughput light stimulation device, the LITOS plate, which enables optogenetic activation across multiple experimental conditions simultaneously (Appendix \ref{app:experiment}) \cite{hohener_litos_2022}. 
To ensure that the measured fluorescence has stabilized and to remove effects from residual signaling dynamics, we include a 4-hour cool-down period after signal termination before measuring $g$. 
This allows Wnt pathway effectors, such as stabilized $\upbeta$-catenin, to return to baseline levels \cite{rosen_anti-resonance_2025}. Since $g$ is not degraded and cell-division in the cool-down period is negligible, the value of $g$ represents a robust measure for gene expression as a consequence of the Wnt pulse.

Histograms of $g$ for a given a signal duration $t$ are left-skewed, long-tailed, and unimodal (see Fig. \ref{fig:stdev_and_mean_tf}B for signal durations $t=5$, $10$, $15$, and $20$ hours). 
We observe that the histograms are well-described by gamma distributions, and that the mean response $\mu_g(t)$ is directly proportional to the standard deviation $\sigma_g(t)$ (Fig. \ref{fig:stdev_and_mean_tf}C). The latter implies that the gamma distribution is parametrized by a constant \textit{shape} parameter $k$ and a time-dependent \textit{scale} parameter $\theta(t)$:
\begin{equation}\label{eq:likelihood}
    p(g \vert t) = \frac{1}{\Gamma(k) \theta (t)^k} g^{k-1} \, e^{-g/\theta (t)},
\end{equation}
where $\Gamma$ is a gamma function. The mean and variance of $g$ are given by:
\begin{align}
\mu_{g}(t) &= k \theta(t), \\
\sigma_g^2(t) &= k \theta(t)^2,
\end{align}
respectively. 
The distribution of Eq. \ref{eq:likelihood} predicts that one can collapse all histograms by normalizing each by their standard deviation (Fig. \ref{fig:stdev_and_mean_tf}D): indeed, after rescaling, all data collapses onto a gamma distribution with shape parameter $\hat{k} = 2.88 \pm 0.01$ and unit variance.

To identify how the scale parameter $\theta(t)$ depends on the Wnt signal duration $t$, we plot the mean $\mu_g(t)$ for all experimental conditions (Fig. \ref{fig:stdev_and_mean_tf}E). For Wnt signals longer than ${\sim} 1$ hour, we find that the mean gene expression grows linearly with the signal duration. Therefore, the scale parameter $\theta(t) \approx a t$ must also grow linearly with time, where we estimate $\hat{a} = 23.0 \pm 0.1 \text{ a.u. hr}^{-1}$. We add a small correction term $\epsilon e^{-t/\tau}$
to the scale parameter $\theta(t)$ to fit the data in the regime $t \lesssim \tau \approx 1$ hour (Appendix \ref{app:gamma_distributions}).

We note that the gamma distribution was fit empirically and is thus a phenomenological description of the data. We choose it here due to its convenient parameterization. Other long-tailed distributions, such as log-normal, may also fit the data well.
These long-tailed distributions, as well as the linear relationship between the standard deviation and the mean, have been suggested to hold universally for protein fluctuations among cells in a population \cite{salman_universal_2012} and emerge in bottom-up stochastic models of gene expression  \cite{koch_logarithm_1966, friedman_linking_2006, cai_stochastic_2006}.

Next, we quantify how precisely we can reconstruct the Wnt signal from the gene expression. Given the broad, long-tailed distributions with substantial overlap between experimental conditions, we anticipate that the information-transmission in the pathway will appear limited.

\section{Inferring the Wnt signal from gene expression in single cells}\label{sec:inference}

We can view our system analogously to a communication channel $t \rightarrow g$ with transmission probability $p(g\vert t)$ (Fig. \ref{fig:stdev_and_mean_tf}F). 
As such, we quantify how much information about the input $t$ is captured by the output $g$ using the \textit{mutual information} \cite{cover_elements_2012, shannon_mathematical_1948}:
\begin{equation}
    I(g;t) = \int_0^\infty \mathrm{d} t \int_0^\infty \mathrm{d} g \, p(g \vert t)\, p(t)\, \mathrm{log}_2 \bigg( \frac{p(g \vert t)}{p(g)} \bigg).
\end{equation}
This mutual information $I(g;t)$ captures (in \textit{bits}) how much we expect to learn about the Wnt signal by observing the gene expression. For the rest of this manuscript, we use the phrase ``input signal" or ``Wnt signal" to refer to the Wnt signal duration $t$.

The mutual information requires knowledge of the distribution of input signals $p(t)$, also referred to as the \textit{prior distribution} \cite{shannon_mathematical_1948,cover_elements_2012,bialek_biophysics_2012}. 
A sensible prior distribution which favors no particular signal condition, like the experiment, is one that is uniform over all available signals $t \in [0,\infty)$.
For this uniform prior, we obtain $ I(g;t) \approx 0.67 \text{ bits}$. Since 1 bit is the minimum required to reliably distinguish two states (e.g. an ``on-off" switch), this result suggests that the gene expression carries less than the information required to support even a binary regulatory decision. 

The numerical value of the mutual information can be difficult to assess abstractly. It is bounded from above by the entropy of the input distribution, which in turn depends on the size of the state space of possible input signals. For example, if the input distribution includes several discrete states, a mutual information of 1 bit does not necessarily imply that any two particular states are neatly distinguishable. Therefore, it can be useful to employ quantities other than the mutual information that allow us to more clearly identify which signals become confused in the information transmission from input to output. 

\begin{figure}[ht]
    \includegraphics[width=0.44\textwidth]{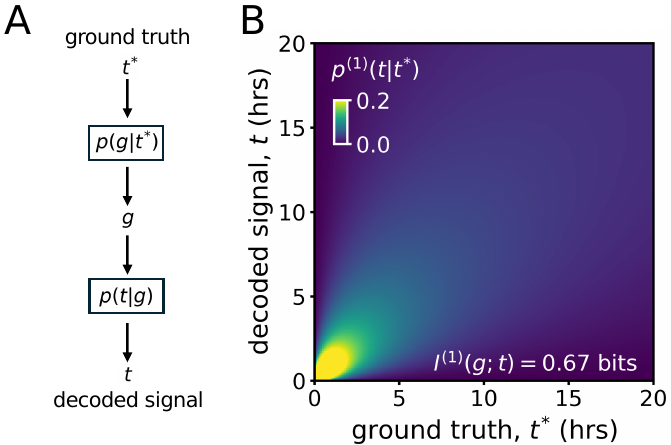}
    \caption{(A) The input signal $t^{*}$ leads to a gene output $g$ drawn from the transmission probability $p(g \vert t^{*})$. Based on a measurement of $g$, one can use the posterior distribution $p(t \vert g)$ to infer the input signal. (B) Decoding map $p^{(1)}(t \vert t^{*}) $ from Eq. \ref{eq:decoding_map_1cell}, showing the average probability assigned to $t$ by the posterior $p(t\vert g)$ given that the true signal is $t^{*}$. Here, the input distribution $p(t)$ is uniform over all possible signals $t \in [0,\infty)$ hours.}\label{fig:decodingmap_1cell}
\end{figure}

To do so, we ask how well one can infer the optogenetic Wnt signal $t$ from a measurement of the gene expression $g$. This is captured by the posterior distribution $p(t \vert g)$, which one obtains from Bayes' theorem,
$p(t \vert g) = p(g \vert t) \, p(t) / p(g)$
where $p(g) = \int_0^\infty \mathrm{d} t \, p(g\vert t) p(t) $. 
With a uniform distribution $p(t)$ over the interval $t \in [0,\infty)$, we find 
\begin{equation}\label{eq:posterior_1}
     p(t \vert g) = \frac{p(g\vert t)}{\int_0^\infty \mathrm{d} t' \, p(g \vert t')} \approx a(k-1) \, p(g \vert t).
\end{equation}
where the final expression is valid in the regime $t \gtrsim 1$ hour. In principle, features of the posterior $p(t \vert g)$ can be used to quantify the precision in this inference problem. Decoding errors, such as the variance of inferred $t$ around its true value, are often used to quantify inference precision \cite{zdeborova_statistical_2016}. However, such metrics rely on selecting a decoding rule, such as the posterior mean or MAP estimate, which may be misleading when the posterior $p(t\vert g)$ is skewed, heavy-tailed, or multi-modal \cite{Petkova_Tkačik_Bialek_Wieschaus_Gregor_2019}. Here, we use a \textit{decoding map} to quantify our ability to decode without subscribing to an estimator. Decoding maps have been used to quantify positional precision from gap gene expression patterns in the early fly embryo \cite{Petkova_Tkačik_Bialek_Wieschaus_Gregor_2019, Bauer_Petkova_Gregor_Wieschaus_Bialek_2021}.

The decoding map quantifies the average posterior $p(t \vert g)$ generated from a true input $t^{*}$.  
To construct the decoding map, we consider a Markov chain in Fig. \ref{fig:decodingmap_1cell}A, and integrate out the regulatory output through which we intend to infer:
\begin{equation}\label{eq:decoding_map_1}
    p^{(1)}( t \vert t^{*}) = \int_{0}^\infty \mathrm{d} g \, p( t  \vert  g) \, p(g  \vert t^{*}).
\end{equation}
The superscript ``$(1)$" refers to the fact that we are considering gene expression from a single ($N=1$) cell. 
While the benefit of decoding maps is most obvious for multi-dimensional $g$, where they provide a means to visualize the precision in the inference in a two dimensional object, they can also be useful for scalar $g$: we will use them later to visualize the performance of different signal encodings.
If the gene expression provides enough information to reconstruct the Wnt signal accurately, the density $p^{(1)}( t \vert t^{*})$ will be sharply peaked around the diagonal $t=t^{*}$.  

We can compute the distribution $p^{(1)}( t \vert t^{*})$ analytically in the regime $t,\,t^{*} \gtrsim 1$ hour, by inserting the posterior from Eq. \ref{eq:posterior_1} into Eq. \ref{eq:decoding_map_1} and performing the change of variables $g' = g (1/\theta(t) + 1/\theta(t^{*}))$,
to obtain:
\begin{equation}\label{eq:decoding_map_1cell}
p^{(1)}(t \vert t^*) \approx \frac{\Gamma(2k-1)}{\Gamma(k)\, \Gamma(k-1)} \frac{ (t t^{*})^{k-1}}{ (t+t^{*})^{2 k-1}}.
\end{equation}
The distribution in Eq. \ref{eq:decoding_map_1cell} is a beta-prime distribution, and the normalizing constant can be identified as a beta-function $B(k, k-1) = \Gamma(k) \, \Gamma(k-1)/\Gamma(2k-1)$ \cite{abramowitz_handbook_1965}.
We plot the decoding map in Fig. \ref{fig:decodingmap_1cell}B, and observe that the width of the decoding map broadens linearly with the Wnt signal duration $t^{*}$. This suggests that absolute decoding errors also grow proportional to $t^{*}$, while relative errors remain constant.

Next, we ask how reliable information transfer from Wnt to a single target-gene could be possible. To do so, we note that the mutual information between the Wnt signal and gene expression depends not only on the channel $p(g \vert t)$, which we take as given from the experimental data, but also on how one chooses the input signals $p(t)$.

\section{Optimal encoding of Wnt signals uses a discrete distribution}\label{sec:opt_encoding}

\begin{figure}
    \centering
    \includegraphics[width=0.40\textwidth]{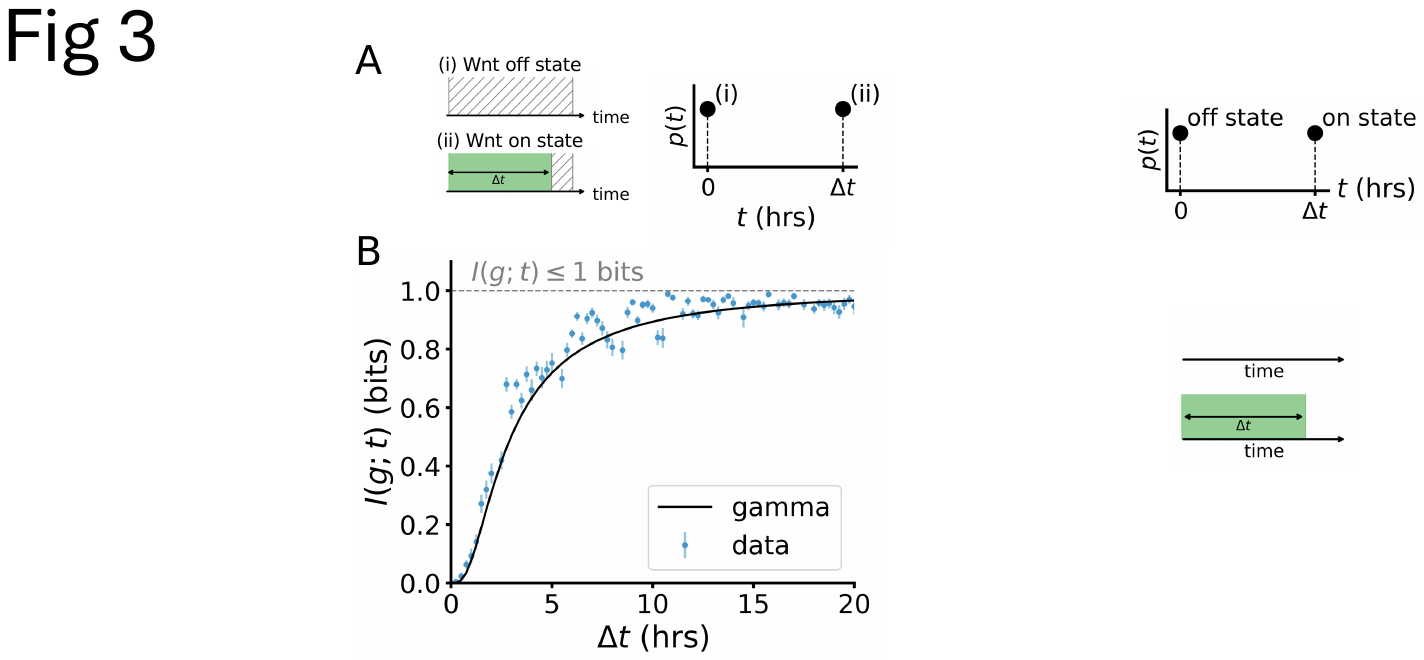}
    \caption{(A) Binary input distribution of optogenetic Wnt signals, containing (i) an ``off" state of $t=0$ hours and (ii) an ``on" state of $t=\Delta t$ hours. (B) Mutual information $I(g;t)$ as a function of the duration $\Delta t$ of the on state. Results from the data (blue) and predictions from the gamma distribution in Eq. \ref{eq:likelihood} (black) are shown. Error bars obtained via subsampling. The upper bound $I \le 1$ (gray line) corresponds to perfect distinguishability of the on and off states}.\label{fig:binary prior}
\end{figure}

As a first step towards optimizing the signal distribution, we focus on a binary input distribution consisting of a Wnt ``off" state ($t=0$ hours) and a single Wnt ``on" state of duration $t=\Delta t$ hours (Fig. \ref{fig:binary prior}A). We explore this setup, since for noisy channels with limited capacity (on the order of 1 bit or less), an efficient coding strategy is to use two maximally distinguishable signal states \cite{smith_information_1971, cover_elements_2012, levchenko2014cellular, Tkacik_Callan_Bialek_2008, mijatovic2025}.  
We find that an optogenetic Wnt signal of approximately $\Delta t \approx 10$ hours is necessary to reliably distinguish the ``on" state from the ``off" state, approaching the information-theoretic upper bound $I(g;t) \leq 1$ bits.

This finding is of biological interest: cells may use Wnt signaling to make binary cell-fate decisions, for example between remaining undifferentiated or committing to mesoderm \cite{kicheva2023control}. In this context, our results suggest that such a binary decision is only reliable if the Wnt ``on" signal persists for durations longer than ${\sim} 10$ hours. This timescale is biologically realistic and lies well within the doubling time of the cells (ca. 20--30 hours \cite{bairoch_cellosaurus_2018, yang_large-scale_2019, moosemiller_hekcell_split}).
In addition, recent work in the intestinal crypt has shown that stem cells commit to differentiation 
after Wnt signaling is lost for ca. 10 hours \cite{zon_loss_2024}. This observation suggests that a binary encoding scheme, based on the sustained presence or absence of Wnt signaling, could have biological relevance.

\begin{figure}
    \centering
    \includegraphics[width=0.48\textwidth]{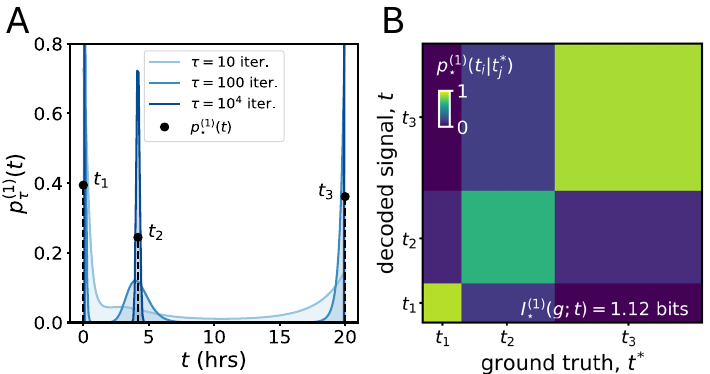}
    \caption{(A) Optimal encoding of optogenetic Wnt signals for single cells: the Blahut--Arimoto algorithm converges to a discrete solution $p_{\star}^{(1)}(t)$, consisting of three optimally distinguishable Wnt signal durations or ``symbols". (B) Decoding map $p_{\star}^{(1)}(t \vert t^\ast)$ obtained using the optimal prior: at the cost of discretizing the space of input signals, we gain distinguishability (cf. Fig. \ref{fig:decodingmap_1cell}B).}\label{fig:BA_convergence}
\end{figure}

Next, we optimize the signal distribution $p(t)$ to obtain the maximally achievable mutual information or \textit{channel capacity}: 
\begin{equation}\label{eq:channel_capacity}
    I_\star = \max_{p(t)} I(g; t).
\end{equation}
The capacity-achieving distribution $p_\star (t)$ tells us how to encode Wnt signals to create maximally distinguishable gene expression outcomes within the noisy constraints. 
In most cases, this optimization is analytically intractable. Instead, we optimize numerically using the Blahut--Arimoto (BA) algorithm \cite{blahut_computation_1972,arimoto_algorithm_1972}. The algorithm converges to a discrete solution (Fig. \ref{fig:BA_convergence}A): the optimal encoding of optogenetic Wnt signals selects a set of three discrete signals (or ``symbols") at $t_1$, $t_2$, and $t_3$. We obtain a capacity of 
\begin{equation} I_{\star}^{(1)} \approx 1.12 \text{ bits,}
\end{equation}
which is a significant improvement over the naive uniform encoding. 

Convergence of the BA algorithm to the discrete solution $p_{\star}(t)$ is slow compared to the convergence to the information capacity $I_\star$, especially if the density of symbols is high. We can exploit the knowledge that $p_{\star}(t)$ is discrete to significantly accelerate convergence to the optimal solution \cite{dauwels_numerical_2005, mattingly_maximizing_2018,abbott_scaling_2019}. To initialize the distribution, we use a weighted sum of $K$ delta-functions, representing $K$ discrete symbols:
\begin{equation}\label{eq:disc_prior}
    p_{\star}(t) = \sum_{i=1}^K w_i \delta(t - t_i).
\end{equation}
We iteratively optimize their locations $t_i$ using gradient descent, while updating the weights $w_i$ using a BA-type update rule. To find the optimal $K$, we use lower and upper bounds to the information capacity to either add or remove a delta-function after convergence (Appendix \ref{ap:algorithms}). 

It is interesting to note that under noisy conditions, non-trivial inputs beyond binary ``on/off'' encoding may be optimal for cells using Wnt signaling.
The decoding map in Fig. \ref{fig:BA_convergence}B visualizes the way the optimal encoding improves information transmission: even though the input signals are not perfectly distinguishable based on their outputs, information transmission is improved compared to a binary input distribution containing two (almost) completely distinguishable states.

\section{Decoding from \textit{N} outputs}\label{sec:opt_encodingN}

\begin{figure*}
    \includegraphics[width=0.99\textwidth]{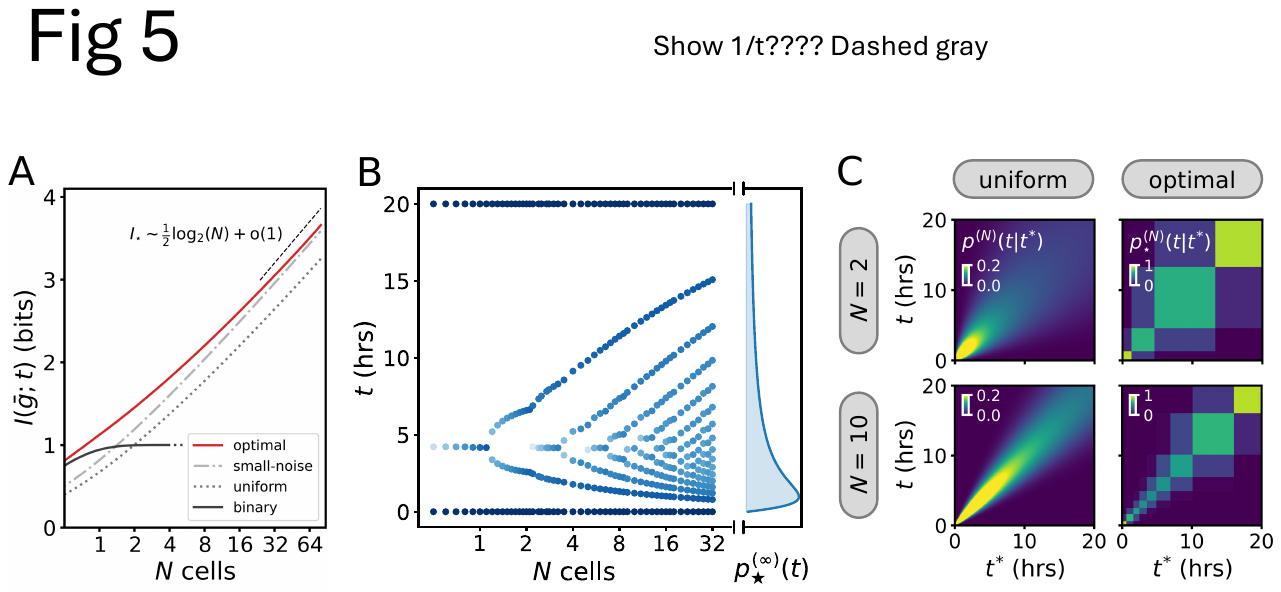}
    \caption{(A) We show the information capacity achieved by the optimal encoding of Wnt signals (red line), and show convergence to analytical results in the small-noise and large-noise regimes. (B) The optimal code for Wnt signaling consists of a discrete number of symbols (blue dots). As the effective noise decreases, the optimal number of symbols increases, and approaches a continuous optimal code $p^{(\infty)}_{\star}(t)$ with a heavy tail that decays as $\sim 1/t$. The color of the markers (blue shade) indicates the relative probability mass of the symbols. (C) Decoding maps visualize how encoding strategies affect signal inference. Shown are decoding maps $p^{(N)}(t \vert t^{*})$ for ensembles of $N=2$ (top row) and $N=10$ (bottom row) cells. A uniform prior over opto-Wnt signals (left column) leads to broad posterior distributions, while the optimized discrete choice of signals (right column) yields more distinguishable responses and higher mutual information $I^{(N)}_{\star}(\bar{g} ; t)$, at the cost of discretizing the space of input signals.}\label{fig:capacity_dots_hessian}
\end{figure*}

We pursued the above analysis under the assumption that the probability with which a particular cell expresses a target gene $g$ is set by the observed experimental response of TopFlash. However, the distribution of TopFlash observed in our cell culture may be noisier than a biologically relevant output of cells in realistic tissue settings \cite{chakrabarti2020circadian,sandberg2005molecular, liu2021self}. 
Therefore, we ask in this section how our analysis changes if the biologically relevant output is systematically more precise than the TopFlash distributions we observe. For example, cells might communicate output with neighbors via surface signaling or molecular exchange, 
or have access to multiple regulatory targets. If the observed TopFlash distribution is representative also of other such outputs,
we can think of these extra outputs as additional samples from the same distribution.

We consider $N$ gene expression outputs $\boldsymbol{g} = (g_1, g_2, \dots, g_N)$, where each output $g_i$ is sampled from a gamma distribution, and use $\boldsymbol{g}$ to decode the Wnt signal $t$. This can correspond to, for example, cells producing $N$ regulatory outputs; alternatively, cells may have access to the mean output of $N$ neighbors, $\bar{g} = \sum_{i=1}^N g_i / N$, through cell-cell communication. Mathematically, it turns out that both options are equivalent. If the responses $g_i$ are independent, one can show that the sample mean $\bar{g}$ is a \textit{sufficient statistic}, implying that all information contained in $\boldsymbol{g}$ about the input signal $t$ is preserved in $\bar{g}$ (Appendix \ref{app:sufficient_stats}). Indeed, for the HEK293T cells in our experiment, we find negligible correlation between TopFlash expression of neighboring cells (Appendix \ref{app:spatial}). Hence, the mutual information satisfies $I(\bar{g}; t) = I(\boldsymbol{g}; t)$ and decoding is identical $p(t \vert \bar{g}) = p(t \vert \boldsymbol{g})$.
Therefore, we can consider the sample mean $\bar{g}$ of $N$ cells in what follows.

Since $\bar{g}$ is the mean of $N$ identically gamma-distributed random variables with shape parameter $k$ and scale parameter $\theta(t)$, its distribution is also gamma, with shape parameter $Nk$ and scale parameter $\theta(t)/N$. The distribution of $\bar{g}$ given a Wnt signal $t$ is thus:
\begin{equation}\label{eq:group_likelihood}
    p(\bar{g} \vert t) = \frac{1}{\Gamma(Nk) (\theta (t)/N)^{Nk}} \bar{g}^{Nk-1} e^{-N\bar{g}/\theta (t)}.
\end{equation}
The mean and variance are given by $\mu_{\bar{g}}(t) = k \theta(t)$ and $\sigma^2_{\bar{g}}(t) = k \theta(t)^2 / N$, respectively; as such, $N$ in Eq. \ref{eq:group_likelihood}  not only represents an integer number of samples, but can also be seen as a continuous parameter that changes the effective noise by a factor of $1/\sqrt{N}$.

We now ask how the mutual information and decoding change as we increase $N$. We can expect the mutual information to scale asymptotically as $I \sim (1/2) \, \mathrm{log}_2 N$ (Appendix \ref{app:smalllargenoise}). We confirm this scaling for both a prior distribution $p(t)$ that is uniform as well as one that is optimized (Fig. \ref{fig:capacity_dots_hessian}A). The mutual information using the binary prior, where our input is restricted to two signal durations, is limited by definition to 1 bit and is not a good choice for maximizing the mutual information when the effective map from input to output becomes more precise. However, in regimes where the noise is large ($N \lesssim 1$), a binary encoding comes close to achieving the information capacity.

In the limit of large $N$, or small effective noise, one can derive an analytic expression for the optimal $p(t)$ \cite{tkacik_information_2008, tkacik_information_2011, bialek_biophysics_2012,walczak_optimizing_2010,dubuis_positional_2013,Bauer_Petkova_Gregor_Wieschaus_Bialek_2021, bauer2022,bauer_information_2023}. 
In this small-noise approximation, we assume that $p(\bar{g} \vert t)$ is a narrow Gaussian distribution, and that we can calculate $p(t \vert \bar{g})$ by performing an expansion (Appendix \ref{app:smalllargenoise}); then, taking a variational derivative of $I(\bar{g};t)$ with respect to $p(t)$, one finds:
\begin{equation}\label{eq:small_noise_prior}
    p^{(\infty)}_{\star}(t) \propto \frac{1}{\sigma_{\bar{g}}(t)} \bigg \vert \frac{\mathrm{d} \mu_{\bar{g}}(t)}{\mathrm{d} t} \bigg \vert.
\end{equation}
Since both $\mu_{\bar{g}}(t)$ and $\sigma_{\bar{g}}(t)$ grow linearly with $t$ for longer Wnt signals, it follows from Eq. \ref{eq:small_noise_prior} that the tail of $p^{(\infty)}_{\star}(t)$ decays as $\sim 1/t$.

As expected, the small-noise approximation approaches the information capacity from below (Fig. \ref{fig:capacity_dots_hessian}A) and is a good approximation for $N \gtrsim 20$. Notably, we can show that this optimal continuous encoding from the small-noise limit in Eq. \ref{eq:small_noise_prior} is equivalent to the \textit{Jeffreys prior} (Appendix \ref{app:smalllargenoise}). The Jeffreys prior is a non-informative prior that is invariant to changes in parametrization, defined as $p_\text{J}(t) \propto \vert \mathcal{I}(t) \vert^{1/2}$, where $\mathcal{I}(t)$ is the \textit{Fisher information} \cite{jeffreys_invariant_1997}:
\begin{equation}
    \mathcal{I}(t) = \int_0^\infty \mathrm{d} \bar{g} \ p(\bar{g}\vert t) \bigg( \frac{\partial \, \mathrm{log} \, p(\bar{g} \vert t)}{\partial t} \bigg)^2.
\end{equation}
Indeed, it is known that in the limit of an infinite number of identical, independent trials of the same experiment (i.e. $N \rightarrow \infty$), the prior that maximizes the mutual information between input and output converges weakly to the Jeffreys prior \cite{scholl_shannon_1998}.

Next, we investigate how the numerically optimized prior $p^{(N)}_\star(t)$ changes as $N$ increases. Since we know that the optimal prior consists of three discrete symbols for $N=1$ and should approach the continuous distribution in Eq. \ref{eq:small_noise_prior} for large $N$, we expect that it will admit an increasing number of symbols as $N$ increases. We find that this is indeed the case:
Fig. \ref{fig:capacity_dots_hessian}B shows a bifurcation-like diagram of the positions and weights of the optimal prior distribution, where symbols split into two and additional symbols are added as $N$ increases. 
For high $N$, the density of the symbols starts approaching the optimal distribution $p^{(\infty)}_{\star}(t)$ from the small-noise approximation. 

As before, we visualize how the optimization improves decoding performance and the mutual information with decoding maps (Fig. \ref{fig:capacity_dots_hessian}C). Unlike the uniform prior, which leads to smoothly narrowing posteriors as $N$ increases (approaching a Gaussian for large $N$), the optimized prior increases the mutual information by admitting more discrete symbols. The optimal number of symbols $K$ in the optimal prior $p_{\star}^{(N)}(t)$ follows an asymptotic scaling law $I_{\star} \sim (3/4) \, \mathrm{log}_2 K$, consistent with recent literature (Appendix \ref{app:scaling_law}) \cite{mattingly_maximizing_2018,abbott_scaling_2019}. 
The discretization enables better distinguishability between inputs, as illustrated by increased activity along the diagonal of the decoding map.
At the same time, the optimal prior does not achieve perfect distinguishability of symbols, reflected by the remaining off-diagonal activity. The optimal input distribution, like in the $N=1$ case, finds a balance between distinguishability and adding additional symbols. An alternative strategy is to encode fewer symbols $K'<K$ than optimal, and approach the bound $\mathrm{log}_2 K’$ bits: this leads to better distinguishability but does not achieve the maximum mutual information.

The fact that the optimal input distribution is discrete 
is interesting: there may be biological situations in which cells want to distinguish between a discrete number of cell fates, such as the different germ layers.
In practice, cell fates are regulated by complex networks of multiple input signals and genetic targets, a more intricate situation than the one we study here.  
It is therefore not clear if our optimal input distribution carries direct biological meaning. Yet, it is conceivable that also for these more complex input spaces, discrete inputs are optimal and it is therefore interesting to investigate if cells attempt to map signals they receive to discrete input states.
We also note that the optimal input distribution is one that permits slight errors, which suggests that some uncertainty in inference of signals is inherently part of optimal information-processing; as such, some observed cellular noise in differentiation could be part of an information-maximization strategy.

\begin{figure}
\includegraphics[width=0.48\textwidth]{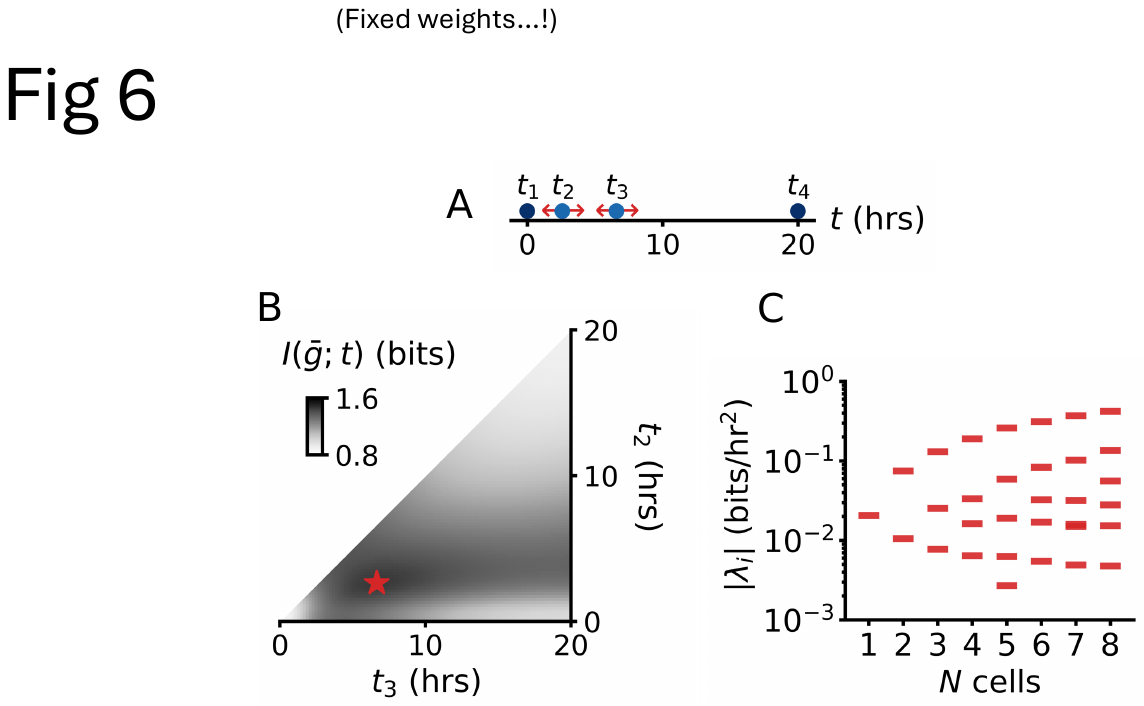}
    \caption{The exact duration of individual symbols does not need to be fine-tuned. (A) For $N=2$, the optimal prior consists of $K=4$ symbols (cf. Fig. \ref{fig:capacity_dots_hessian}B). We vary the positions of two symbols $t_2$ and $t_3$, while keeping their weights fixed. (B) Information $I(\bar{g};t)$ (colorshade), shows a broad optimum (red star) as a function of the position of peaks $t_2$ and $t_3$. (C) The Hessian matrix (Eq. \ref{eq:hessian}) has a sloppy spectrum that widens as $N$ increases: symbols at longer durations become more sloppy and symbols at shorter durations become more stiff.}\label{fig:hessian}
\end{figure}

We observe that the numerical optimization for the optimal prior distribution converges more quickly to the correct value of the mutual information than to the correct number of delta-functions $K$ and their positions $t_i$, especially as $N$ becomes larger \cite{mattingly_maximizing_2018}. 
This implies that the information-landscape at the optimum is smooth, and has some directions where parameters for the prior distribution still change while the optimum is almost attained. These directions are typically referred to as ``sloppy" directions \cite{gutenkunst,transtrum2015perspective} and their presence has important implications for the ability of biological systems to show variability in parameter space, even at the optimum \cite{ bauer_optimization_2025}. 
Indeed, in Fig. \ref{fig:hessian}A we show the mutual information for $N=2$ as a function of the positions of two out of four delta-functions and observe a broad optimum with different sensitivities depending on the direction one moves away from the optimum.  

The sloppiness can be quantified using the Hessian matrix of the cost-function, in this case the mutual information $I(\bar{g};t)$ for a given $N$ \cite{Bauer_Petkova_Gregor_Wieschaus_Bialek_2021,bauer_optimization_2025}. Calculating this Hessian can be numerically difficult. Here, we have access to the functional form of the probability distribution $p(g \vert t)$, and can therefore calculate the Hessian with respect to the positions $t_i$ of the discrete symbols in the optimal encoding $p_{\star}^{(N)}(t)$ (Appendix \ref{app:hessian}). Writing $p_{\star}^{(N)}(t)$ as in Eq. \ref{eq:disc_prior}, the Hessian matrix becomes:
\begin{align}\label{eq:hessian}
\chi_{ij} &= \frac{\partial^2 I(\bar{g}; t)}{\partial t_{i} \partial t_{j}} \\
&= \frac{w_i}{\mathrm{log} \, 2} \int_0^\infty \mathrm{d} \bar{g}\, \bigg\{\delta_{ij} \bigg[\frac{\partial^2 p(\bar{g} \vert t_i)}{\partial t_i^2} \mathrm{log} \bigg(\frac{p(\bar{g}\vert t_i)}{p(\bar{g})} \bigg) \nonumber \\
&\quad+ \frac{1}{p(\bar{g} \vert t_i)} \bigg( \frac{\partial p(\bar{g} \vert t_i)}{\partial t_i}\bigg)^2 \bigg] - \frac{w_j}{p(\bar{g})}  \frac{\partial p(\bar{g} \vert t_i)}{\partial t_i} \frac{\partial p(\bar{g} \vert t_j)}{\partial t_j}\bigg\}.\nonumber
\end{align}
The eigenvectors of $\chi$ determine directions in parameter space $t_i$ that have independent effects on the mutual information, and the eigenvalues $\lambda_i$ tell us the sensitivity along these directions. 
We evaluate the Hessian at the stationary point that maximizes $I(\bar{g};t)$. After diagonalization, we indeed observe a sloppy spectrum (Fig. \ref{fig:hessian}B), with eigenvalues spanning ca. 2 decades. 
The most stiff eigendirections correspond to the shorter durations, where the density of symbols is highest.
As $N$ increases, the spectrum broadens: symbols at longer durations becoming more sloppy, and those at shorter durations become more stiff.
Practically, the fact that the optimal prior is sloppy implies that the optimal signal encoding does not need to be fine-tuned \cite{Bauer_Petkova_Gregor_Wieschaus_Bialek_2021, bauer_optimization_2025}; this could indeed be one advantage of information transmission using channels with similarly long-tailed distributions of gene expression outputs.

In deriving Eq. \ref{eq:hessian} we assumed weights $w_i$ are fixed; alternatively, one can keep the weights optimized while varying the position of the symbols. In that case we also obtain a sloppy spectrum (Appendix \ref{app:hessian}). We emphasize that Eq. \ref{eq:hessian} (and the extension for variable weights in Appendix \ref{app:hessian}) are general and can be used to obtain the sensitivity spectrum of a discrete prior for any choice of transmission probability.

\section{Discussion}
In this work, we optimized information transmission from optogenetic Wnt signals to (representative) Wnt target gene expression. We measured expression of a fluorescent reporter in a HEK cell line for different optogenetic signal durations, and found that gamma distributions with a constant shape parameter describe the output distributions well across different input conditions. We calculated the distribution of input signals that maximizes information transfer and found that it consists of a discrete set of signals that allows for optimal decoding from the output. We also explored how this optimal set of signals would change if cells had access to multiple instances of the output we measure: for example, because they average multiple genetic outputs or exchange outputs with neighboring cells. As the effective noise decreases, the optimal input distribution evolves from three discrete symbols to a continuous distribution: we showed that the latter can be obtained from the small noise approximation or, equivalently, the Jeffreys prior. Using decoding maps, we visualized how the optimal choice of of optogenetic inputs improves signal inference, and finally, we showed that the optimal signals do not need to be fine-tuned. In the following, we discuss the biological and theoretical impact of our findings.

\paragraph{Discrete Wnt input signals?} 
The canonical Wnt signaling pathway is active in different types of cells, often with the goal to direct mammalian cells towards differentiation.
The effective features of Wnt signals that cells respond to depend on context and may include absolute or relative concentrations, thresholds, timers, and combinatorial interactions with other signaling pathways. Here, we focus on the Wnt signal duration as a potential input, since it can be precisely controlled in our optogenetic setup.
Durations of Wnt signals may have developmental relevance, particularly in the context of organoid development \cite{zon_loss_2024, spence_directed_2011, sato_growing_2013, van_den_brink_single-cell_2020}.
We found that two Wnt input signals (a binary prior) provide more information than a uniform distribution when cells have to decide based on a single output: when the Wnt input is either no signal or a signal lasting at least ca. 10 hours, the output expression is precise enough to reliably distinguish between the two.
Such a binary response between two states, e.g. cell differentiation or remaining in the undifferentiated state, could be relevant for Wnt biologically \cite{Amel2023,van2014symmetry}.
This timescale of ca. 10 hours could also be biologically reasonable: it fits well within a cell's division cycle, and connects to recent findings in gut cells, where differentiation is triggered only after Wnt signal loss lasting about 10 hours, while shorter transient losses leave cells undifferentiated \cite{zon_loss_2024}.

Our work to infer optimal priors would imply, in analogy with ideas with efficient coding, that cells either experience strongly different Wnt signals, or map the Wnt signals they experience onto signals that are strongly differentiable. 
This idea is similar to ideas from optimal parameter inference with limited data \cite{mattingly_maximizing_2018}. While it is unclear if this happens biologically, it is interesting to observe that both recent work on differentiation timers \cite{zon_loss_2024}, as well as clustering of cells with similar inputs \cite{repina2023optogenetic}, could suggest that cells indeed try to map the input they receive onto clear binary states (here, the presence or absence of Wnt). In our analysis, we find that cells could, in principle, resolve a third, intermediate state as well. We do not know whether this additional state is used as such \textit{in vivo}, but it seems possible that multiple Wnt input states can be decoded in some contexts \cite{rosenbloom2020beta,mcnamara2024recording} and 
it is interesting that this possibility emerges purely from the distributions observed in the experiment and a model-free optimization framework.

In addition to suggestions for what signals cells may receive naturally, our work also has meaning from an engineering perspective: the optimal optogenetic signals we infer could be used with the goal to engineer signals that populations of cells will be able to respond to precisely. This could be interesting also in the context of synthetic development. 

\paragraph{Optimal priors for systematically lower noise, connections to population coding, and decoding maps} 
Our work connects to previous work on optimizing information transmission in gene regulation. That a binary prior (or switch) can optimize transmission for noisy systems has also previously been discussed in the context of genetic networks \cite{Tkacik_Callan_Bialek_2008, walczak_optimizing_2010, Bauer_Petkova_Gregor_Wieschaus_Bialek_2021, mijatovic2025}. An interesting and perhaps more surprising finding of our work is the increasing number of discrete signals as the effective noise in the signaling pathway decreases: 
While the precise phase diagram of symbols depends on the underlying distribution, in general, additional symbols emerge as the noise decreases since the Shannon-optimal prior balances complexity with the ability to resolve different input signals. 
Ref. \cite{mattingly_maximizing_2018} showed this effect in the context of Bayesian inference, to predict what models should be inferred for noisy data; in our work, it is either the cell or the engineer that would employ the optimal priors for reliable signal processing. The fact that there is sloppiness of the exact positions of these signals can make near-optimal information transmission easier than expected.

We investigated signaling via multiple cells or outputs as a way  to improve information transmission; this is equivalent to using multiple communication channels and is related to population coding in neural systems \cite{averbeck_neural_2006, pillow_spatio-temporal_2008, schneidman_synergy_2003}.
An alternative approach to consider if information transmission from signals to outputs is higher than observed is to consider repeated stimuli on single cells: indeed, recent work shows that single cells responding to repeated opto-genetic stimuli show a more reproducible output---higher mutual information---compared to population-level measurements \cite{kramar_single_2025}.

To visualize how the encoding of input signals influences inference, we use decoding maps alongside the mutual information. They are reminiscent of recurrence plots for non-linear dynamical systems, showing at what times a dynamical system reverts to a state it has visited before. Recurrence plots can be viewed as a measure of the ``predictability" of the system \cite{marwan_recurrence_2007}, similar to how decoding maps can visualize ambiguity in inference from multi-modal posterior distributions \cite{Petkova_Tkačik_Bialek_Wieschaus_Gregor_2019,Bauer_Petkova_Gregor_Wieschaus_Bialek_2021}. In our case, the decoding maps visualize how the optimal encoding discretizes the input space to improve distinguishability between signals, and how signal inference becomes more precise as the effective noise decreases. 

\paragraph{Outlook and future work} 

Our work is directly applicable to gene expression responses that follow gamma distributions, and we expect qualitatively similar results for other long-tailed distributions that are common observed for gene expression \cite{salman_universal_2012,koch_logarithm_1966, friedman_linking_2006, cai_stochastic_2006}.
This suggests that the implications of our work may go beyond Wnt signaling.
Our systematic reduction of noise would also predict for those signaling pathways that discrete input distributions maximize information flow. While we do not know if these discrete intermediate durations are a relevant input \textit{in vivo}, Wnt/$\upbeta$-catenin systems exhibit multi-state dynamic decoding in other contexts, and developing embryoid systems pass through intermediate Wnt activity states during patterning \cite{rosenbloom2020beta,mcnamara2024recording}. 
In the context of general gene regulatory responses, it seems possible that smooth inputs are mapped onto a discrete but not infinite number of outputs, such as a finite number of segments in the fly embryo body plan: in this context, it may be beneficial to map signal inputs onto different discrete states. A more detailed investigation of evidence of the use of discrete signal priors in gene regulatory or other biological contexts is an interesting direction for future work.

Our work relates closely to the problem of selecting effective models that maximize the information extracted from finite data, as discussed in Ref. \cite{mattingly_maximizing_2018}. In their framework, this corresponds to choosing an Bayesian prior that maximizes the mutual information between parameters and predictions.  
In our system, there are two concrete applications: first, the optimal prior can predict signal durations that should be used in synthetic, engineered experiments if cells are to respond distinguishably. Second, if cells operate consistent with the model selection work of  Ref. \cite{mattingly_maximizing_2018}, the cell itself should map the signal onto a discrete state space to optimally extract information. It will be interesting to explore the optimal input distributions in multi-input settings \cite{kicheva2023control, camacho2024combinatorial,lehr2025self} in the future. 

\begin{acknowledgments}
We thank Florian Berger, William Bialek, Aneta Koseska,  Pieter Rein ten Wolde, and the members of the Bauer group for useful discussions. We acknowledge funding from the NWO VIDI Talent Programme, NWO/VI.Vidi.223.169 (MB). This collaboration was started during a workshop at the Aspen Center for Physics, which is supported by National Science Foundation grant PHY-2210452.
\end{acknowledgments}

\appendix

\section{Experimental methods}\label{app:experiment}

Human 293T cells were cultured at 37°C and 5\% CO2 in Dulbecco’s Modified Eagle Medium, high glucose GlutaMAX (Thermo Fisher Scientific, 10566016) medium supplemented with 10\% fetal bovine serum (Atlas Biologicals, F-0500-D) and 1\% penicillin-streptomycin (Thermo Fisher Cat. No. 15140-122). Clonal 293Ts containing CRISPR tdmRuby3-$\upbeta$-cat, oLRP6-Puro (AddGene ID: 249712) and pPig-8X-TOPFlash-tdIRFP-Puro (AddGene ID: 249713) were obtained from Dr. Ryan Lach's previous experiments \cite{lach_nucleation_2022, rosen_anti-resonance_2025}.

We seeded these cells onto a 96-well glass bottom plate (Cellvis Cat. No.: P96-1.5H-N) coated with fibronectin (Thermo Fisher Cat. No.: 33010018) in Dulbecco’s Modified Eagle Medium, high glucose GlutaMAX (Thermo Fisher Scientific, 10566016) medium supplemented with 10\% fetal bovine serum (Atlas Biologicals, F-0500-D) and 1\% penicillin-streptomycin. Cells were allowed to adhere to the plate for 24 hours. Afterwards, cells were stimulated via a benchtop LED array purpose-built for light delivery to cells in standard tissue culture plates (LITOS) \cite{hohener_litos_2022}. Light was patterned to cover the entire surface of intended wells of plates used, rather than a single microscope imaging field. Post-LITOS stimulation, cells were moved into dark conditions for 4 hours to ensure the Wnt pathway was properly deactivated prior to imaging. After 4 hours, cells were then imaged using a Nikon W2 SoRa spinning-disk confocal microscope equipped with an incubation chamber maintaining cells at 37°C and 5\% CO2. 
We then used our previously published image analysis pipeline (Methods, Image Analysis from \cite{rosen_anti-resonance_2025}). We quantified the TopFlash-tdiRFP fluorescent intensity (measured in arbitrary units, a.u.) by taking the median pixel value for each individual segmented cell; we refer to this quantity as $g$.

We analyzed $g$ for populations of ca. $1500 \pm 800$ cells that were optically stimulated for up to 24 hours, which is of order of the doubling time
of 293T cells (20--30 hours \cite{bairoch_cellosaurus_2018, yang_large-scale_2019, moosemiller_hekcell_split}). We included data of up to until 20 hours in this experiment, since 20 hours is on the lower end of possible cell cycle estimates, and we observe a clear linear increase of mean TopFlash expression with time in this time frame. After 1-3 division cycles, cells grow beyond the volume of the plate, and tracking becomes more difficult as this time frame is approached. Therefore, a time frame of ca. 20 hours marks both a natural end to our experiment and lies well within the regime of reliable experimental data acquisition. Finally, we note that cell division is negligible during the ca. 4-hour cool-down window after the Wnt signal, since only ca. 15\% of cells are expected to divide in this time frame. We therefore expect no decay of TopFlash intensity in this regime, and TopFlash accumulation has been found to be stable in similar experiments even for over 10 hours after the end of the Wnt ``on" condition; for details on TopFlash dynamics, we refer to Ref. \cite{rosen_anti-resonance_2025}. Analysis from the smaller dataset of Ref. \cite{rosen_anti-resonance_2025} (not shown) suggests that TopFlash distributions at longer cool-down windows are also gamma-distributed with the same shape parameter $k$ but different scale parameters; since most analytic results in our manuscript depend explicitly on $k$, our results do not sensitively depend on the cool-down window and could easily be extended to others.

\section{Fitting the gene expression data}\label{app:gamma_distributions}

Gene expression distributions observed in cell cultures are frequently long-tailed, and have been successfully modeled using e.g. gamma, negative binomial, or lognormal distributions \cite{salman_universal_2012}. These choices are not only empirically well-matched to the data, but also mechanistically plausible in systems where gene expression is shaped by multiple interacting timescales \cite{koch_logarithm_1966, friedman_linking_2006, cai_stochastic_2006}. In our case, we find that our distributions are particularly well described by a gamma distribution:
\begin{equation}\label{eq:likelihood_app}
p(g  \vert  k, \theta(t)) = \frac{1}{\Gamma(k) \theta(t)^k}g^{k-1} e^{-g/\theta(t)},
\end{equation}
parametrized by a constant shape parameter $k$ and a time-dependent scale parameter $\theta(t)$. In fact, one can show that this parametrization is our only choice given that the mean $\mu_g(t)$ and standard deviation $\sigma_g(t)$ are directly proportional (Fig. \ref{fig:stdev_and_mean_tf}C). Our aim here is to estimate $k$ and $\theta(t)$ from the data using a maximum-likelihood estimate. 

We denote the data for TopFlash as $g_{ij}$ and the signal durations as $t_i$, where $i=1, \dots, n$ and $j = 1, \dots, m_i$. The $i$th experimental condition is populated by $m_i$ different cells at the end of the experiment. We first consider the regime where $\theta(t) \approx at$. Thus, our task is to estimate $k$ and $a$. The likelihood function is given by: 
\begin{equation}
\mathcal{L}(k,a) = \prod_{i=1}^n \prod_{j=1}^{m_i}  p (g_{ij}  \vert  k, a, t_i),
\end{equation}
and hence our log-likelihood is
\begin{equation}
\begin{split}
\log( \mathcal{L}(k,a)) &= \sum_{i=1}^n \sum_{j=1}^{m_i} \log ( p (g_{ij} \vert  k, a, t_i) ) \\ 
&=  \sum_{i=1}^n \sum_{j=1}^{m_i} \bigg[ -\log( \Gamma(k) ) - k\log(a) - k \log (t_i) \\
&\quad + (k-1) \log(g_{ij}) -\frac{g_{ij}}{a t_i} \bigg].
\end{split}
\end{equation}
Setting
\begin{equation}
\frac{\partial \log (\mathcal{L}) }{ \partial a } \bigg{\vert}_{\hat{k},\, \hat{a}} = 0,
\end{equation}
we obtain:
\begin{equation}\label{eq:ahatestimator}
\hat{a} = \frac{1}{\hat{k}} \left( \frac{1}{\sum_{l=1}^n m_l} \sum_{i=1}^n \sum_{j=1}^{m_i} \frac{g_{ij}}{t_i} \right).
\end{equation}

Further, setting 
\begin{equation}
\frac{\partial \log (\mathcal{L}) }{ \partial k } \bigg{\vert}_{\hat{k},\, \hat{a}} = 0,
\end{equation}
we get:
\begin{equation}
\begin{split}
\psi^{(0)}(\hat{k}) &= \log ( \hat{k}) +  \frac{1}{\sum_{l=1}^n m_l} \sum_{i=1}^n \sum_{j=1}^{m_i}  \bigg[- \log (g_{ij}) \\ &\quad + \log \bigg( \frac{1}{\sum_{r=1}^n m_r} \sum_{p=1}^n \sum_{q=1}^{m_p} \frac{g_{pq}}{t_p} \bigg)  + \log(t_i) \bigg],
\end{split}
\end{equation}
where $\psi^{(0)}(k) = \mathrm{d} \log (\Gamma(k)) / \mathrm{d} k $ is the polygamma function of order $0$. We can solve this numerically for $\hat{k}$ without too much trouble, though we can continue analytically to very good approximation using the asymptotic expansion:
\begin{equation}
\psi^{(0)}(k) \sim \log (k) - \sum_{l=1}^\infty \frac{B_l}{l k^l}.
\end{equation}
Here $B_l$ are Bernouilli numbers with the convention $B_1 = +\frac{1}{2}$. Keeping the first two terms in the sum, we obtain the following estimator for $k$:
\begin{equation}\label{eq:khatestimator}
\hat{k} = \frac{1 + \sqrt{1+ \frac{4}{3} C }}{4 C },
\end{equation}
where
\begin{equation}
\begin{split}
C &= \frac{1}{\sum_{l=1}^n m_l} \sum_{i=1}^n \sum_{j=1}^{m_i}  \bigg[- \log (g_{ij}) \\ &\quad + \log \bigg( \frac{1}{\sum_{r=1}^n m_r} \sum_{p=1}^n \sum_{q=1}^{m_p} \frac{g_{pq}}{t_p} \bigg)  + \log(t_i) \bigg].
\end{split}
\end{equation}

We obtain, using Eqs. \ref{eq:ahatestimator} and \ref{eq:khatestimator}:
\begin{align}
\hat{k} &= 2.88 \pm 0.01, \\
\hat{a} &= 23.0 \pm 0.1 \text{ hr}^{-1}.
\end{align}
where the error is obtained using the asymptotic normality of maximum-likelihood estimators.

If we take the scale parameter to be $\theta(t) = at$, the gamma distribution in Eq. \ref{eq:likelihood_app} has a singularity at $t=0$. The singularity is not physical, as it implies that any signal $t>0$ is perfectly distinguishable from $t=0$, and more importantly does not match the experimental data in this regime. To regularize the behavior of $p(g \vert t)$ near $t=0$, we add a small exponential term to the scale parameter:
\begin{align}
    \theta(t) &= a (t + \epsilon e^{-t/\tau}),
\end{align}
where $\hat{\epsilon} = 0.86$ hours and $\hat{\tau} = 0.87$ hours are positive constants which we estimate from the data. The mean and variance are therefore given by:
\begin{align}
    \mu_{g}(t) &= ka (t + \epsilon e^{-t/\tau}), \label{eq:mean_corr}\\
    \sigma_{g}^2(t) &= k a^2(t + \epsilon e^{-t/\tau} )^2,\label{eq:var_corr}
\end{align}
respectively. At times $t\gtrsim \tau$ the exponential term becomes irrelevant and we recover $\theta(t) \approx at$.

\section{Algorithm for computing the channel capacity}\label{ap:algorithms}

We want to maximize the mutual information $I(g ; t)$ with respect to the input distribution $p(t)$:
\begin{equation}\label{eq:channel_capacity_app}
    I_\star = \max_{p(t)} I(g;t),
\end{equation}
where $I_\star$ is the channel capacity and $p(g \vert t)$ is fixed. In most cases, this optimization is analytically intractable and we must proceed numerically. The Blahut--Arimoto (BA) algorithm is the standard algorithm for solving this problem \cite{blahut_computation_1972,arimoto_algorithm_1972}. 
One starts with an initial guess $p^{(0)}(t)$ for the input distribution, and with each iteration it is updated as follows:
\begin{equation}
    p^{(\tau)}(t) = \frac{1}{Z^{(\tau-1)}} p^{(\tau-1)}(t) \, e^{f^{(\tau -1)}_{\mathrm{KL}}(t)},
\end{equation}
where
\begin{equation}
    f^{(\tau)}_{\mathrm{KL}}(t) = \int_0^\infty \mathrm{d} g \, p(g \vert t) \, \mathrm{log} \bigg( \frac{p(g \vert t)}{p^{(\tau)}(g)}\bigg),
\end{equation}
and
\begin{equation}
    p^{(\tau)}(g) = \int_{0}^{T} \mathrm{d} t \, p(g \vert t) \, p^{(\tau)}(t).
\end{equation}
Practically, we restrict ourselves to a finite domain $t\in [0,T]$, where $T$ is the maximum signal duration.

After $\tau$ iterations, the lower bound to the channel capacity is given by:
\begin{equation}
    I_L^{(\tau)} = \frac{1}{\mathrm{log} \, 2}\int_{0}^T \mathrm{d} t \, p^{(\tau)}(t) \, f^{(\tau)}_{\text{KL}}(t),
\end{equation}
and an upper bound is given by:
\begin{equation}
    I_U^{(\tau)} = \max_{t} \frac{f^{(\tau)}_{\text{KL}}(t)}{\mathrm{log} \, 2}.
\end{equation}
As such, we iterate until convergence:
\begin{equation}
    I_U^{(\tau)} - I_L^{(\tau)} < \epsilon.
\end{equation}
and use $I_\star \approx I_L^{(\tau)}$ and $p_{\star}(t) \approx p^{(\tau)}(t)$ as our estimates for the channel capacity and the optimal input distribution, respectively. The optimization problem in Eq. \ref{eq:channel_capacity_app} is convex and guaranteed to converge to the global maximum \cite{cover_elements_2012}. 

As shown in Fig. \ref{fig:BA_convergence} in the main text and as noted by Mattingly et al. \cite{mattingly_maximizing_2018}, convergence to a discrete solution in the interior of the domain $t \in [0,T]$ is rather slow compared to the boundaries, especially when there is high density of delta functions. To overcome this, we can exploit the knowledge that $p_{\star}(t)$ is discrete by starting with a sum of $K$ delta functions:
\begin{equation}
    p(t) = \sum_{i=1}^K w_i \delta(t - t_i),
\end{equation}
and adjusting the weights $w_i$ and positions $t_i$ iteratively. Starting with $K$ equally spaced delta functions, we use the BA algorithm to adjust the weights $w_i$ until convergence. After this, we use the gradient:
\begin{equation}\label{eq:info_gradient}
    \frac{\partial I}{\partial t_i} = \frac{w_i}{\mathrm{log}\,2} \int_0^\infty \mathrm{d} g \, \frac{\partial p(g \vert t_i)}{\partial t_i} \, \mathrm{log} \bigg( \frac{p(g \vert t_i)}{p(g)}\bigg),
\end{equation}
to adjust the positions $t_i$. Iterating the adjustment of the weights via the BA algorithm and the positions by gradient ascent, we can converge to the optimal discrete solution.

In contrast to the BA algorithm, the optimization over $w_i$ and $t_i$ is not convex: in particular, it depends on the number of peaks $K$ we define beforehand. After convergence, we can compute $f_{\text{KL}}(t)$ everywhere in the domain. If $\max_t f_{\text{KL}}(t)$ is greater than $f_{\text{KL}}(t)$ evaluated at any of the peaks $t_i$, we have to add another delta function \cite{mattingly_maximizing_2018}. This way, we ensure that (i) we have converged to the global maximum and (ii) that we have used the optimal number of delta functions. 

\section{Sufficient statistics for independent, identical, gamma-distributed variables}\label{app:sufficient_stats}

We find that there is negligible spatial correlation in the gene expression $g$ (Appendix \ref{app:spatial}). Hence, we can treat the cells as responding independently conditional on the Wnt signal $t$. Below, we show that when one considers a group of $N$ cells, the arithmetic mean $\bar{g}$ is a sufficient statistic for the signal duration $t$. That is, $\bar{g}$ contains as much information about $t$ as the whole dataset $\boldsymbol{g}$, as claimed in the main text.

A single cell $i$ in a group of $N$ cells exposed to a Wnt signal of duration $t$ responds independently by expressing output $g_i \sim \text{Gamma}(k, \theta(t))$. Hence, the likelihood function for the group of $N$ cells is given by:
\begin{align}\label{eq:likelihood_group1}
        p(\boldsymbol{g} \vert t) &= \prod_{i=1}^N p(g_i \vert t), \\
        \label{eq:likelihood_group2}
        &= \frac{\prod_{i=1}^N g_i^{k-1} }{\Gamma(k)^N \theta (t)^{Nk}} e^{-\sum_{i=1}^N g_i /\theta(t)}, 
\end{align}
In Sec. \ref{sec:opt_encodingN}, we claim that the arithmetic mean $\bar{g} \equiv \frac{1}{N} \sum_{i=1}^N g_i$ is a sufficient statistic for $t$. A quick way to see this is by observing that Eq. \ref{eq:likelihood_group2} satisfies Fisher-Neyman factorization \cite{halmos_application_1949}. That is, it can be written in the form $p(\boldsymbol{g} \vert t) = h(\boldsymbol{g}) \, f(\bar{g},t)$ for nonnegative functions $h$ and $f$, from which the defining property $p(\boldsymbol{g}\vert \bar{g}, t) = p(\boldsymbol{g} \vert \bar{g})$ follows. Below, however, we derive this property explicitly.

The likelihood of $\bar{g}$ conditioned on $t$ can be written as:
\begin{equation}\label{eq:likelihood_gbar}
    p(\bar{g} \vert t) = \frac{1}{\Gamma(Nk) (\theta(t)/N)^{Nk}} \bar{g}^{Nk-1} e^{-N\bar{g}/\theta(t)},
\end{equation}
which follows from the addition of $N$ independent and gamma-distributed random variables. Using Bayes' theorem, we obtain:
\begin{equation}\label{eq:likelihood_g_given_gbar_t}
    p(\boldsymbol{g} \vert \bar{g}, t) = \frac{p(\boldsymbol{g} \vert t)}{p(\bar{g} \vert t)}.
\end{equation}
Substituting Eq. \ref{eq:likelihood_group2} and Eq. \ref{eq:likelihood_gbar} into the above, we get:
\begin{equation}\label{eq:likelihood_g_given_gbar_t2}
    p(\boldsymbol{g} \vert \bar{g}, t) = \frac{\prod_{i=1}^N g_i^{k-1}}{\bar{g}^{Nk-1}}\frac{ \Gamma(Nk) } { (N^k \Gamma(k))^N } = p(\boldsymbol{g} \vert \bar{g}),
\end{equation}
as required. The last equality follows as all $t$-dependence has been canceled out.

Sufficiency of $\bar{g}$ also implies the posterior distributions $p(t \vert \boldsymbol{g})$ and $p(t \vert \bar{g})$ are identical. This can be seen by applying Bayes' theorem
$
    p(t \vert \boldsymbol{g}) = p(\boldsymbol{g} \vert t) \, p(t)/p(\boldsymbol{g}).
$
Substituting $p(\boldsymbol{g} \vert t) = p(\boldsymbol{g} \vert \bar{g}) \, p(\bar{g} \vert t)$ and $p(\boldsymbol{g}) = p(\boldsymbol{g} \vert \bar{g})\, p(\bar{g})$, we obtain:
\begin{equation}
    p(t \vert \boldsymbol{g}) = \frac{ p(\bar{g} \vert t) p(t)}{p(\bar{g})} = p(t \vert \bar{g}),
\end{equation}
as required.

We can also show that the mutual information satisfies $I(t ; \boldsymbol{g}) = I(t ; \bar{g})$. This makes precise the statement that the statistic $\bar{g}$ contains as much information about the signal $t$ as the whole dataset $\boldsymbol{g}$. To show this, we use the data-processing inequality for the mutual information in two ways. Firstly, as $\bar{g}$ is a function of the dataset $\boldsymbol{g}$, we must have that:
\begin{equation}
    I(t ; \boldsymbol{g}) \geq I(t ; \bar{g}).
\end{equation}
Secondly, we have just shown that $p(\boldsymbol{g} \vert \bar{g}, t) = p(\boldsymbol{g} \vert \bar{g})$. This implies that we have a Markov chain $t \rightarrow \bar{g} \rightarrow \boldsymbol{g}$, to which we can also apply the data-processing inequality:
\begin{equation}
    I(t ; \boldsymbol{g}) \leq I(t ; \bar{g}).
\end{equation}
Together, these inequalities imply that $I(t ; \boldsymbol{g}) = I(t ; \bar{g})$, as required.

Finally, we can show that the decoding maps $p_{\boldsymbol{g}}(t \vert t^{*})$ and $p_{\bar{g}}(t \vert t^{*})$ are identical. Starting with $p_{\boldsymbol{g}}(t \vert t^{*})$, we can introduce an integral over $\bar{g}$ using the law of total probability:
\begin{align}
    p_{\boldsymbol{g}}(t \vert t^{*}) &= \int_{0}^\infty \mathrm{d} \boldsymbol{g} \, p( t  \vert  \boldsymbol{g}) \, p(\boldsymbol{g}  \vert t^{*}), \\
    &= \int_{0}^\infty \mathrm{d} \boldsymbol{g} \int_{0}^\infty \mathrm{d} \bar{g} \, p( t  \vert  \boldsymbol{g}) \, p(\boldsymbol{g}  \vert \bar{g}) \, p(\bar{g} \vert t^{*}).
\end{align}
By changing the order of integration and using the fact that $p(t \vert \boldsymbol{g}) = p(t \vert \bar{g})$, we obtain:
\begin{align}
    p_{\boldsymbol{g}}(t \vert t^{*}) &= \int_{0}^\infty \mathrm{d} \bar{g} \, p( t  \vert  \bar{g}) \, p(\bar{g} \vert t^{*}) \int_{0}^\infty \mathrm{d} \boldsymbol{g} \, p(\boldsymbol{g}  \vert \bar{g}), \\
    &= p_{\bar{g}}(t \vert t^{*}).
\end{align}
As such, the decoding maps are identical whether we consider the gene expression in the whole group $\boldsymbol{g}$ or just the mean expression $\bar{g}$.

\section{Negligible spatial correlations in TopFlash expression}\label{app:spatial}

\begin{figure}[ht]
    \centering
    \includegraphics[width=0.48\textwidth]{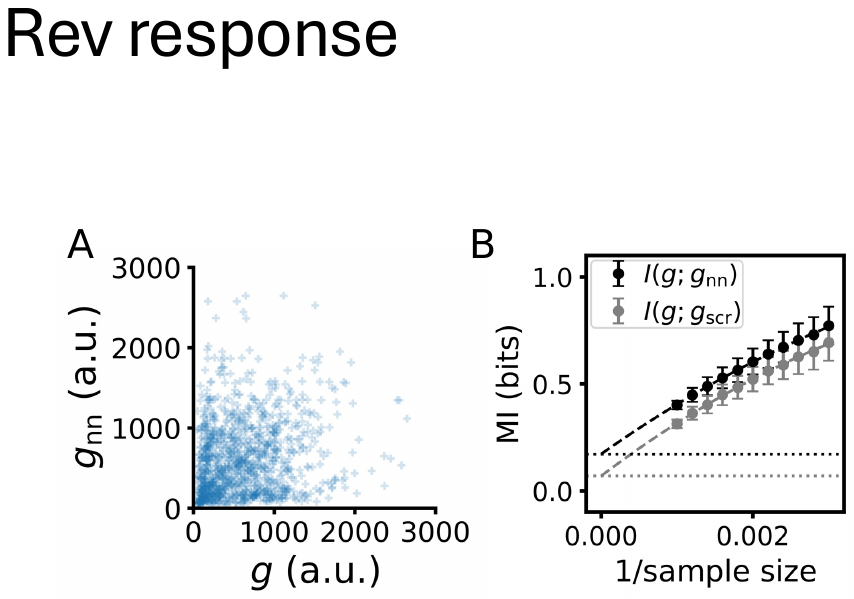}
    \caption{There are negligible spatial correlations between responses $g$ to the optogenetic Wnt signal. Here, we show data from a repeat experiment for $t = 12.5$ hours. (A) Scatterplot showing TopFlash expression $g$ versus the expression of the nearest neighboring cell, $g_\text{nn}$. (B) Mutual information $I(g;g_\text{nn})$ and $I(g;g_\text{scr})$, where $g_\text{scr}$ is a scrambled permutation of $g$, as a function of inverse sample size. Extrapolating to infinite sample size we obtain $I(g;g_\text{nn}) \approx 0.2 \text{ bits}$ and $I(g;g_\text{scr}) \approx 0.1 \text{ bits}$, which are negligible compared to the joint entropies $H(g, g_\text{nn}) \approx 21.0 \text{ bits}$ and $H(g, g_\text{scr}) \approx 21.1 \text{ bits}$.}\label{fig:spatial}
\end{figure}

In the main text, we consider the vector response $\boldsymbol{g}$ of a group of $N$ cells to an optogenetic Wnt signal of duration $t$. We treated each response as independent. To justify this assumption, we investigate here whether there are spatial correlations; specifically, whether the TopFlash response $g$ of a particular cell is correlated to that of its nearest neighbor $g_{\text{nn}}$. Indeed, we find no significant correlation between $g$ and $g_\text{nn}$: the results from a repeat experiment for $t = 12.5$ hours are shown in Fig. \ref{fig:spatial}. The scatterplot between $g$ and $g_\text{nn}$ reveals an uncorrelated cloud (Fig. \ref{fig:spatial}A), and the mutual information $I(g;g_{\text{nn}})$ is negligible (Fig. \ref{fig:spatial}B).

\section{Small-noise regime and equivalence to the Jeffreys prior}\label{app:smalllargenoise}

As before, we seek to find the distribution $p_{\star}(t)$ that maximizes the mutual information $I(\bar{g};t)$. In this section, we consider regimes where the optimization is analytically tractable. In Fig. \ref{fig:capacity_dots_hessian} in the main text, we show that we can analytically compute the channel capacity $I_\star$ in regimes where the effective noise is really small ($N \gg 1$). We derive these results here. 

As the effective noise approaches zero, it is known that the optimal code for a communication channel becomes continuous \cite{mattingly_maximizing_2018,abbott_scaling_2019}. In this regime, we can use the fact that the noise is small to derive $p_{\star}(t)$ analytically. In literature, this is often referred to as the small-noise approximation and is widely used for studying information transmission in biological systems \cite{tkacik_information_2008, tkacik_information_2011,walczak_optimizing_2010,dubuis_positional_2013,Bauer_Petkova_Gregor_Wieschaus_Bialek_2021, bauer2022,bauer_information_2023}. We will also illustrate that the prior obtained in this limit is formally the same as the Jeffreys prior, a non-informative prior that is often used in Bayesian statistics \cite{jeffreys_invariant_1997}.

We explore the gene regulatory response for a particular signal duration, $t \rightarrow \bar{g}$, which occurs with probability
    \begin{equation}
        p(\bar{g} \vert t) = \frac{1}{\Gamma(Nk) (\theta(t) / N)^{Nk}} \bar{g}^{Nk-1} e^{-N\bar{g}/\theta(t)}.
    \end{equation}
In the limit of small noise ($N \gg 1$), we can approximate the gamma distribution as a narrow Gaussian distribution with a mean and variance given by $\mu_{\bar{g}}(t) = k \theta(t)$ and $\sigma_{\bar{g}}^2(t) = k \theta (t)^2 / N$ respectively. 

We write the mutual information as a difference of entropies:
\begin{equation}\label{eq:mutual_information_sn}
    I(\bar{g};t) = H(\bar{g}) - H(\bar{g} \vert t),
\end{equation}
where:
\begin{align}\label{eq:entropy_g_sn}
    H(\bar{g}) = - \int_0^\infty \mathrm{d} \bar{g} \, p(\bar{g}) \, \mathrm{log}_2 \, p(\bar{g}), 
\end{align}
and:
\begin{align}\label{eq:conditional_ent_sn}
    H(\bar{g} \vert t) = - \int_0^T \mathrm{d} t \, p(t) \int_0^\infty \mathrm{d} \bar{g} \, p(\bar{g} \vert t) \, \mathrm{log}_2 \, p(\bar{g} \vert t).
\end{align}
When the noise $\sigma_g(t)$ is small, the mapping $p(\bar{g}\vert t)$ is almost deterministic. As such, we can write:
\begin{equation}
    p(t) \approx p(\bar{g}) \bigg \vert \frac{\mathrm{d} \mu_{\bar{g}}}{\mathrm{d} t} \bigg \vert.
\end{equation}
This allows us to write entropy in Eq. \ref{eq:entropy_g_sn} as:
\begin{equation}
    H(\bar{g}) \approx H(t) + \int_0^T \mathrm{d} t \, p(t) \, \mathrm{log}_2 \bigg \vert \frac{\mathrm{d} \mu_{\bar{g}}}{\mathrm{d} t} \bigg \vert.
\end{equation}
Further, using the Gaussian approximation for $p(\bar{g} \vert t)$, the conditional entropy in Eq. \ref{eq:conditional_ent_sn} becomes:
\begin{equation}
    H(\bar{g} \vert t) \approx \frac{1}{2} \int_0^T \mathrm{d} t \, p(t) \, \mathrm{log}_2 (2 \pi e \sigma_{\bar{g}}^2(t)).
\end{equation}
Having written both $H(\bar{g})$ and $H(\bar{g} \vert t)$ in a way where $p(t)$ is the only ``free'' distribution that we can vary, we can now proceed with the optimization. We add a Lagrangian multiplier to ensure normalization of $p(t)$, and optimize
\begin{equation}
    \mathcal{L}[p(t)] = I(\bar{g} ; t) - \beta \int_0^T \mathrm{d} t \, p(t). 
\end{equation}
Taking the variational derivative with respect to $p(t)$, we get:
\begin{equation}
    \frac{\delta \mathcal{L}}{\delta p(t)} = \mathrm{log}_2 \bigg \vert \frac{\mathrm{d} \mu_{\bar{g}}}{\mathrm{d} t} \bigg \vert - \mathrm{log}_2 \, p(t) - \frac{1}{\mathrm{log} \, 2}  - \frac{1}{2}\mathrm{log}_2 (2 \pi e \sigma_{\bar{g}}^2(t)) - \beta.
\end{equation}
Setting $\delta \mathcal{L} / \delta p(t) = 0$, we obtain:
\begin{equation}\label{eq:optimal_prior_sn}
    p_{\star}(t) = \frac{1}{Z} \frac{1}{\sigma_{\bar{g}}(t)} \bigg \vert \frac{\mathrm{d} \mu_{\bar{g}}}{\mathrm{d} t} \bigg \vert,
\end{equation}
where
\begin{equation}\label{eq:normalizing_constant_sn}
    Z = \int_0^T \frac{\mathrm{d} t}{\sigma_{\bar{g}}(t)} \bigg \vert \frac{\mathrm{d} \mu_{\bar{g}}}{\mathrm{d} t} \bigg \vert,
\end{equation}
is a normalizing constant. Using the optimal input distribution in Eq. \ref{eq:optimal_prior_sn} to evaluate $I(\bar{g};t)$, we obtain the channel capacity given by \cite{bialek_biophysics_2012}:
\begin{equation}\label{eq:channel_capacity_sn}
    I_\star = \mathrm{log}_2 \bigg[ \frac{Z}{\sqrt{2 \pi e}} \bigg].
\end{equation}
Since $\theta(t) \propto t$ for times $t \gtrsim 1$ hour, both $\mu_{\bar{g}}(t)$ and $\sigma_{\bar{g}}(t)$ grow linearly with $t$ in this regime. It follows from Eq. \ref{eq:optimal_prior_sn} that the tail of the optimal prior decays functionally as $\sim 1/t$ (cf. Fig. \ref{fig:capacity_dots_hessian}B).

In Fig. \ref{fig:capacity_dots_hessian}in the main text, we show the mutual information obtained using the prior in Eq. \ref{eq:optimal_prior_sn} derived from the small-noise approximation. Indeed, asymptotically with increasing $N$, we converge to the channel capacity $I_\star$. To derive the functional behavior of the channel capacity $I_{\star}$ with large $N$ we can proceed by inspection. By construction, the mean $\mu_{\bar{g}}(t)$ does not depend on $N$. The variance obeys $\sigma_{\bar{g}}^2(t) \propto 1/N$, and therefore the normalizing constant $Z$ in Eq. \ref{eq:normalizing_constant_sn} satisfies $Z \propto N^{1/2}$. As such, the channel capacity in Eq. \ref{eq:channel_capacity_sn} asymptotically scales as:
\begin{equation}\label{eq:channel_capacity_scaling_sn}
    I_{\star} \sim \frac{1}{2}\mathrm{log}_2 N + \mathrm{o}(1).
\end{equation} 

We also note that the prior obtained in the small-noise limit is formally identical to the Jeffreys prior \cite{jeffreys_invariant_1997}. It is known that in the limit of an infinite number of identically, independent trials of the same experiment (i.e. $N \rightarrow \infty$), the prior that is Shannon-optimal converges weakly to the Jeffreys prior \cite{scholl_shannon_1998}. Indeed, the same is true for a single precise experiment in the limit of small noise. 
We derive the Jeffreys prior below and verify that it is indeed identical to Eq. \ref{eq:optimal_prior_sn}. 

The Jeffreys prior is designed to be invariant under reparametrization of the probability distribution, and is defined as: 
\begin{equation}
    p_\text{J}(t) \propto \vert \mathcal{I}(t) \vert^{1/2},
\end{equation}
where $\mathcal{I}(t)$ is the \textit{Fisher information}:
\begin{equation}\label{eq:fisherinfo}
    \mathcal{I}(t) = \int_0^\infty \mathrm{d} \bar{g} \ p(\bar{g}\vert t) \bigg( \frac{\partial \, \mathrm{log} \, p(\bar{g} \vert t)}{\partial t} \bigg)^2.
\end{equation}
In our case, $p(\bar{g}\vert t)$ is gamma-distributed with shape parameter $Nk$ and a time-dependent scale parameter $\theta(t)/N$:
\begin{equation}
\begin{split}
\mathrm{log} \, p(\bar{g} \vert t) &= - \ \mathrm{log}\, \Gamma(Nk) - Nk\, \mathrm{log} (\theta(t)/ N)  \\
&\quad + (Nk-1) \, \mathrm{log} \, \bar{g} - \frac{N\bar{g}}{\theta(t)}.
\end{split}
\end{equation}
Taking the first derivative and substituting $\mu_{\bar{g}}(t) = k \theta(t)$ and $\sigma^2_{\bar{g}}(t) = k \theta(t)^2/N$, we obtain:
\begin{equation}
\bigg( \frac{\partial \, \mathrm{log} \, p(\bar{g} \vert t)}{\partial t} \bigg)^2 = \frac{1}{\sigma_{\bar{g}}(t)^4} \bigg(\frac{\mathrm{d} \mu_{\bar{g}}(t)}{\mathrm{d} t} \bigg)^2 \big( \bar{g} - \mu_{\bar{g}}(t))^2.
\end{equation}
The expectation over $(\bar{g} - \mu_{\bar{g}}(t))^2$ is just the variance $\sigma_{\bar{g}}^2(t)$; hence, the Fisher information in Eq. \ref{eq:fisherinfo} becomes:
\begin{equation}
    \mathcal{I}(t) = \frac{1}{\sigma_{\bar{g}}(t)^2} \bigg(\frac{\mathrm{d} \mu_{\bar{g}}(t)}{\mathrm{d} t} \bigg)^2.
\end{equation}
The Jeffreys prior can now be written as:
\begin{equation}
    p_\text{J}(t) \propto \frac{1}{\sigma_{\bar{g}}(t)} \bigg \vert \frac{\mathrm{d} \mu_{\bar{g}}}{\mathrm{d} t} \bigg \vert,
\end{equation}
which is identical to the Shannon-optimal prior in the limit of small noise in Eq. \ref{eq:optimal_prior_sn}. We have thus verified that the Jeffreys prior and the prior that optimizes the mutual information are equivalent in the limit of small noise. 

\section{Asymptotic scaling law for the channel capacity}\label{app:scaling_law}

\begin{figure}
    \centering
    \includegraphics[width=0.30\textwidth]{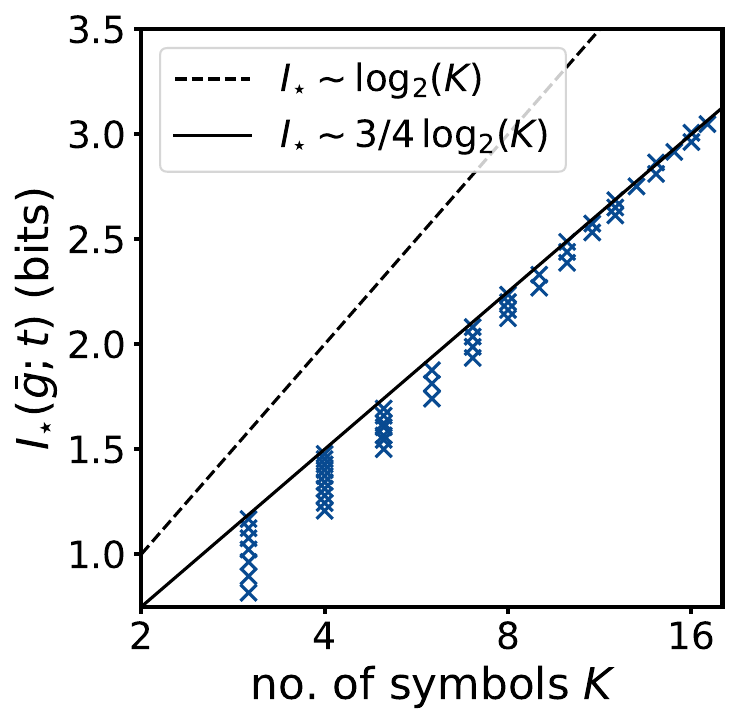}
    \caption{Asymptotic scaling law between the number of symbols $K$ in the optimized prior $p_{\star}(t)$ and the mutual information $I(g;t)$, in agreement with literature \cite{abbott_scaling_2019,mattingly_maximizing_2018}. The bound $I(\bar{g} ; t) \leq \mathrm{log}_2 K$ is shown using a dotted line.}\label{fig:scaling_law}
\end{figure}

In the main text, we explored what happens to our optimal prior $p^{(N)}_{\star}(t)$ as we considered multiple cells $N$. As $N$ becomes larger, the effective noise level decreases---analogous to sending the same message multiple times in a communication channel. As the noise level approaches zero, it is known that the number of symbols $K$ in the optimal code follows an asymptotic scaling law. In this limit, the channel capacity $I_\star$ scales with the logarithm of the number of symbols $K$ as $I_\star \sim (3/4)\, \mathrm{log}_2 K$ \cite{abbott_scaling_2019}. Indeed, we confirm the scaling law for our system in Fig. \ref{fig:scaling_law}, as a non-trivial check of our optimization and to verify consistency with existing literature. The dotted line shows the fundamental limit $I_\star \leq \mathrm{log}_2 K$, which would reach equality if the $K$ symbols were perfectly distinguishable. 

\section{Fine-tuning of the optimal code}\label{app:hessian}
\begin{figure}[ht]
    \centering
    \includegraphics[width=0.48\textwidth]{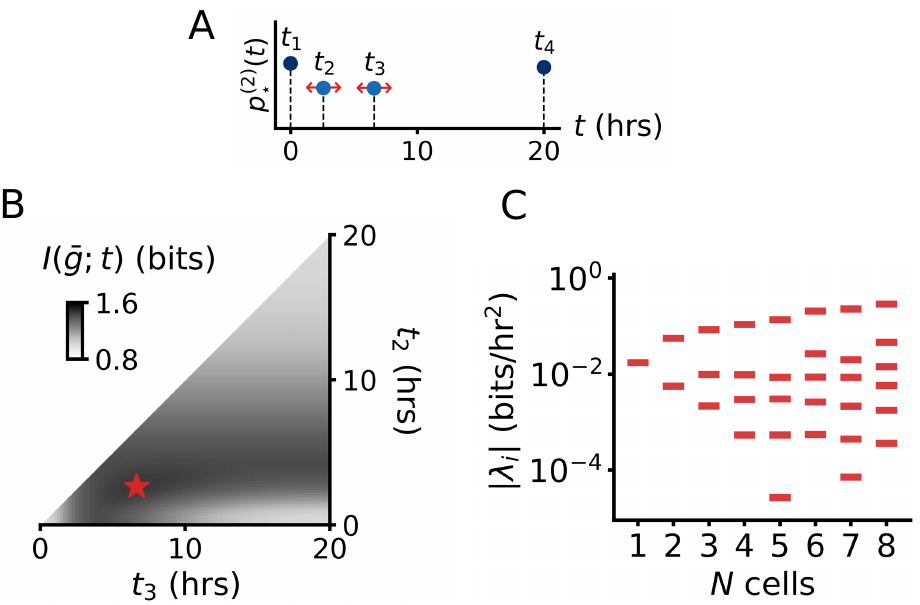}
    \caption{Fine-tuning of symbols in the optimal code; here weights $\{w_i\}$ are kept optimized as positions $\{t_i\}$ are varied. (A) For $N=2$, the optimal prior consists of $K=4$ symbols (cf. Fig. \ref{fig:capacity_dots_hessian}B). We vary the positions of two symbols $t_2$ and $t_3$, while keeping the weights $w_i$ optimized. (B) Information $I(\bar{g};t)$ (colorshade), shows a broad optimum (red star) as a function of the position of peaks $t_2$ and $t_3$. (C) The Hessian matrix (cf. Eq. \ref{eq:hessian_2}) has a sloppy spectrum that widens as $N$ increases.}\label{fig:app_hessian}
\end{figure}

Here we consider a prior distribution $p(t)$ composed of a set of $K$ discrete symbols $\{t_i\}$ with respective weights $\{w_i \}$:
\begin{equation}
p(t) = \sum_{i=1}^K w_i \, \delta(t - t_i),
\end{equation}
and ask to what extent the capacity-achieving distribution $p_\star(t)$ needs to be fine-tuned. Since the BA algorithm converges much more quickly to the channel capacity $I_\star$ than to the optimal prior $p_\star(t)$, we anticipate a sloppy information-landscape around the optimum.

To investigate fine-tuning of the prior, we write the mutual information $I(\bar{g} ;t)$ as a function of the positions $\{t_i\}$ while keeping the weights $\{w_i\}$ fixed:
\begin{equation}
    I(\bar{g};t \vert \{ t_i \} ) = \sum_{i=1}^K  w_i \int_0^{\infty} \mathrm{d} \bar{g} \, p(\bar{g} \vert t_i) \, \mathrm{log}_2 \bigg(\frac{p(\bar{g} \vert t_i)}{p(\bar{g})} \bigg).
\end{equation}
Denoting the optimized positions as $\{t_i^\star \}$, we can expand the mutual information around its maximum, the capacity:
\begin{equation}
    I(\bar{g};t \vert \{ t_i \} ) = I_\star + \frac{1}{2} \sum_{i,j = 1}^K (t_i - t_i^\star) \chi_{ij} (t_j - t_j^\star)  + \dots,
\end{equation}
where $\chi_{ij}$ is the Hessian matrix:
\begin{equation}
    \chi_{ij} = \frac{\partial^2 I}{\partial t_i \partial t_j} \bigg \vert_{\{ t_i^\star \}}.
\end{equation}
The eigenvectors of $\chi$ determine directions in parameter space $\{t_i \}$ that have independent effects on the mutual information, whereas the eigenvalues $\{ \lambda_i \}$ tell us the sensitivity along these directions \cite{Bauer_Petkova_Gregor_Wieschaus_Bialek_2021}. Since we have have fit the data with a functional form for $p(\bar{g} \vert t_i )$, we can proceed analytically to compute the Hessian. 

To proceed, we express the partial derivatives of the marginal distribution $p(g)$ with respect to symbol $t_i$,
\begin{equation}
    \frac{\partial p(\bar{g})}{\partial t_i} = w_i \frac{\partial p(\bar{g} \vert t_i)}{\partial t_i}.
\end{equation}
We can then compute the first derivative of the mutual information with respect to $t_i$: 
\begin{equation}
    \frac{\partial I}{\partial t_i} = \frac{w_i}{\mathrm{log}\, 2} \int_0^\infty \mathrm{d} \bar{g} \, \frac{\partial p(\bar{g} \vert t_i)}{\partial t_i} \, \mathrm{log} \bigg(\frac{p(\bar{g} \vert t_i)}{p(\bar{g})}\bigg).
\end{equation}
Note that this is the same gradient as the one used to iteratively adjust the positions in Eq. \ref{eq:info_gradient}. Taking a second derivative with respect to $t_j$, we obtain:
\begin{equation}\label{eq:hessian_app}
\begin{split}
\chi_{ij} &= \frac{\partial^2 I}{\partial t_{i} \partial t_{j}} \\
&= \frac{w_i}{\mathrm{log} \, 2} \int_0^\infty \mathrm{d} \bar{g}\, \bigg\{\delta_{ij} \bigg[\frac{\partial^2 p(\bar{g} \vert t_i)}{\partial t_i^2} \mathrm{log} \bigg(\frac{p(\bar{g}\vert t_i)}{p(\bar{g})} \bigg) \\
&\quad+ \frac{1}{p(\bar{g} \vert t_i)} \bigg( \frac{\partial p(\bar{g} \vert t_i)}{\partial t_i}\bigg)^2 \bigg] - \frac{w_j}{p(\bar{g})}  \frac{\partial p(\bar{g} \vert t_i)}{\partial t_i} \frac{\partial p(\bar{g} \vert t_j)}{\partial t_j}\bigg\}.
\end{split}
\end{equation}
To compute the integral we insert the functional form $p(\bar{g} \vert t)$ and integrate numerically, after which we can find the spectrum $\{\lambda_i \}$ by diagonalizing $\chi$ (Fig. \ref{fig:capacity_dots_hessian}D). We note that in deriving Eq. \ref{eq:hessian_app} we have not assumed a functional form for $p(\bar{g} \vert t)$, and thus the equation holds for any choice of transmission probability and discrete prior. 

In the above and the main text, we kept the weights $\{w_i\}$ fixed while adjusting the positions $\{t_i\}$ of the symbols. Alternatively, one can keep the weights optimized while varying the position of the symbols. In that case, the Hessian becomes a total derivative $\chi_{ij} = \mathrm{d}^2 I / \mathrm{d} t_i \mathrm{d} t_j$ and receives a contribution from adjusting the weights as one departs from the optimum. Denoting the partial derivatives of the mutual information by $(I_{\boldsymbol{t}\boldsymbol{t}})_{ij} = \partial^2 I / \partial t_i \partial t_j$, $(I_{\boldsymbol{w}\boldsymbol{t}})_{ij} = \partial^2 I / \partial w_i \partial t_j$, and 
$(I_{\boldsymbol{w}\boldsymbol{w}})_{ij} = \partial^2 I / \partial w_i \partial w_j$, we can write the Hessian as:
\begin{equation}\label{eq:hessian_2}
    \chi = I_{\boldsymbol{t} \boldsymbol{t}} - I_{\boldsymbol{w} \boldsymbol{t}}^\mathrm{T} M I_{\boldsymbol{w} \boldsymbol{t}},
\end{equation}
where $M$ is a matrix given by:
\begin{equation}
    M = I_{\boldsymbol{w}\boldsymbol{w}}^{-1} - \frac{I_{\boldsymbol{w}\boldsymbol{w}}^{-1}\boldsymbol{1} \boldsymbol{1} ^\mathrm{T} I_{\boldsymbol{w}\boldsymbol{w}}^{-1}}{\boldsymbol{1}^\mathrm{T} I_{\boldsymbol{w}\boldsymbol{w}}^{-1} \boldsymbol{1}},
\end{equation}
and $\boldsymbol{1} = (1, \dots, 1)$.
Again, these expressions are general and hold for any choice of transmission probability and discrete prior. In Fig. \ref{fig:app_hessian}A we plot the mutual information for $N=2$ as a function of the positions of two out of four delta-functions while keeping the weights optimized. We observe a broad optimum similar to Fig. \ref{fig:hessian}A. In Fig. \ref{fig:app_hessian}B we use Eq. \ref{eq:hessian_2} to plot the spectrum as a function of $N$; spanning four decades, it is more sloppy than the case for fixed weights in the main text.

\bibliography{wnt_inference}

\end{document}